\documentclass[twocolumn]{svjour3}
\smartqed

\usepackage{url}

\usepackage{balance}
\usepackage{amsmath}
\usepackage{pbox}
\usepackage{tabularx}
\usepackage{graphicx}
\usepackage{amsfonts}
\usepackage{rotating}
\usepackage[table,xcdraw]{xcolor}
\usepackage{multirow}
\usepackage{color}
\usepackage{colortbl}
 \usepackage[normalem]{ulem}
 \useunder{\uline}{\ul}{}
\usepackage{hvfloat}
\usepackage{booktabs}
\usepackage{hvfloat}
\usepackage{units}
\emergencystretch=\maxdimen

\usepackage{amssymb}
\usepackage[hyperfootnotes=false]{hyperref}
\usepackage{amssymb}
\usepackage{pifont}
\usepackage{relsize}

\usepackage{pdflscape}
\usepackage{afterpage}

\usepackage{capt-of}
\usepackage{multibib}
\newcites{ONLINE}{Online References}

\usepackage{amssymb}
\usepackage{graphicx}
\usepackage{multirow}
\usepackage{hyperref}
\usepackage{graphics} 
\usepackage{tabularx,booktabs}
\usepackage{times} 
\usepackage{pbox}
\usepackage{rotating}
\usepackage{array}
\usepackage{makecell}
\usepackage{booktabs}

\usepackage{colortbl}

\usepackage[font=small,labelfont=bf,
 justification=justified,
 format=plain]{caption}
\usepackage{amssymb}
\usepackage{pifont}
\usepackage[utf8x]{inputenc}
\usepackage{url}

\usepackage{silence}
\begin{document}

\title{
A Decision Model for Decentralized Autonomous Organization Platform Selection: Three Industry Case Studies


}

\author{Elena Baninemeh \and
Siamak Farshidi \and
Slinger Jansen
}

\institute{
E. Baninemeh \at
Department of Information and Computer Science at Utrecht University, Utrecht, the Netherlands\\
Tel.: +31 6 15 373 513\\
\email{e.baninemeh@uu.nl}
 \and
 S. Farshidi \at
Informatics Institute at University of Amsterdam, Amsterdam, the Netherlands\\
Tel.: +31 6 15 373 513\\
\email{s.farshidi@uva.nl}
 \and
 S. Jansen \at
Department of Information and Computer Science at Utrecht University, Utrecht, the Netherlands\\
Department of Software Engineering at Lappeenranta University of Technology, Lappeenranta, Finland\\
Tel.: +31 6 19 884 880\\
\email{slinger.jansen@uu.nl}
}

\date{Received: date / Accepted: date}

\maketitle

\noindent\textbf{Abstract} 

\noindent\textbf{Context:} 
Decentralized autonomous organizations as a new form of online governance are collections of smart contracts deployed on a blockchain platform that intercede groups of people. A growing number of \textit{Decentralized Autonomous Organization Platforms}, such as Aragon and Colony, have been introduced in the market to facilitate the development process of such organizations. Selecting the best fitting platform is challenging for the organizations, as a significant number of decision criteria, such as popularity, developer availability, governance issues, and consistent documentation of such platforms, should be considered. Additionally, decision-makers at the organizations are not experts in every domain, so they must continuously acquire volatile knowledge regarding such platforms and keep themselves updated. Accordingly, a decision model is required to analyze the decision criteria using systematic identification and evaluation of potential alternative solutions for a development project. 

\noindent\textbf{Method}: We have developed a theoretical framework to assist software engineers with a set of Multi-Criteria Decision-Making problems in software production. This study presents a decision model as a Multi-Criteria Decision-Making problem for the decentralized autonomous organization platform selection problem to capture knowledge regarding such platforms and concepts systematically. 

\noindent\textbf{Results}: We conducted three industry case studies in the context of three decentralized autonomous organizations to evaluate the effectiveness and efficiency of the decision model in assisting decision-makers. The case study participants declared that the decision model provides significantly more insight into their selection process and reduces the cost of the decision-making process. 

\noindent\textbf{Conclusion}: We find that with empirical evidence from the case studies, that decision-makers can make more rational, efficient, and effective decisions with the decision model when they meet their requirements and priorities. Furthermore, the captured reusable knowledge regarding platforms and concepts while building the decision model can be employed by other researchers to develop new concepts and solutions for future challenges.


\section{Introduction}

First Bitcoin in 2008~\cite{nakamoto2008peer}, and later Ethereum in 2014~\cite{wood2014ethereum}, held a powerful promise: decentralized governance, without third party authorization, not just for finance applications such as cryptocurrencies but for any organization. Decentralized Autonomous Organizations (DAOs) were expected to fulfill such promises, enabling people to organize online, relying on blockchain-based systems and smart contracts, and automating their governance~\cite{valiente2020evaluating}.

No widely accepted definition for DAO exists yet. For instance, Buterin~\cite{buterin2014next} explains DAOs as a way to explore those new organizations’ governance rules that could be automated and transparently embedded in a blockchain.  Alternatively, Dhillon et al.~\cite{dhillon2017hyperledger} define it as a blockchain entity built on a consensus of decisions by its members, and Beck et al.~\cite{beck2018governance} define it as a decentralized, transparent, and secure system for operation and governance among independent participants which can run autonomously. This study uses the following statements to refer to a decentralized organization: a DAO offers services or resources to third parties or even hires people to perform specific tasks. Hence, individuals can transact with a DAO in order to access its service or get paid for their contributions~\cite{P2PModels:online}. DAOs are virtual, decentralized entities such as corporations and institutions running entirely autonomously and decentralized on a distributed ledger~\cite{singh2019blockchain}.

A variety of DAO platforms have recently emerged to facilitate the deployment of DAOs in the blockchain by significantly reducing the technological knowledge required and providing DAO software as a service. These DAO platforms enable users with insufficient knowledge on how blockchain works to create a DAO using a template that typically can be customized~\citeONLINE{AragonDA79:online}. A significant number of DAO platforms such as Aragon~\footnote{\url{https://aragon.org/}}, DAOstack\footnote{\url{https://daostack.io/}}, and Colony\footnote{\url{https://colony.io//}} with a broad list of features and criteria are available in the market. This study focuses on these particular DAOs. \textit{The main challenge is selecting the best fitting platform from an extensive set of alternatives based on an extensive list of project requirements.} Besides, for the domain experts, the DAO is typically not their expertise, and they have limited time for acquiring the needed knowledge. 

Technology selection problems in the software production domain can be modeled as a Multi-Criteria Decision-Making (MCDM) problem that deals with evaluating a set of alternatives and consider a set of decision criteria~\cite{farshidi2020multi}. Recently, we introduced a technology selection framework~\cite{farshidi2018decision} that is used to build decision models for MCDM problems and assist decision-makers at software-producing organizations with their decision-making processes~\cite{farshidi2018decision}. Additionally, we have designed and implemented a Decision Support System~(DSS)~\cite{farshidi2018DSS,FarshidiToolPaper2020} for supporting decision-makers with their MCDM problems in software production. The DSS provides a decision model studio for building decision models based on the technology selection framework. Such decision models can be uploaded to the knowledge base of the DSS to facilitate the decision-making process for software-producing organizations according to their requirements and preferences. The DSS provides a discussion and negotiation platform to enable decision-makers at software-producing organizations to make group decisions. Furthermore, the DSS can be used over the full life-cycle and can co-evolve its advice based on evolving requirements. 

In this study, the DAO platform selection process is modeled as an MCDM problem, and the technology selection framework is employed to build a decision model for this MCDM problem. In order to evaluate the efficiency and effectiveness of the decision model in assisting decentralized autonomous organizations, three real-world case studies have been conducted. 

The structure of this article is as follows: Section~\ref{Background} describes layers of the blockchain technology stack and positions the DAO platform selection problem among other blockchain technology selection problems in this domain. Section~\ref{Research Approach} formulates the DAO platform selection problem as an MCDM problem, defines the research questions of the study, and explains our research method, which is based on the design science, expert interviews, document analysis, and case studies. This study reports on the following contributions: 

\begin{itemize}

\item \textbf{Section~\ref{MCDM}} elaborates on how we have mapped the explicit knowledge of DAO experts to the explicit knowledge of DAO platforms that we have captured based on an extensive literature study. 

\item\textbf{Section~\ref{CaseStudies}} explains our empirical observations in the context of three real-world case studies that have been conducted to evaluate the effectiveness and usefulness of the decision model in addressing the  DAO Platform selection problem.

\item\textbf{Section~\ref{AnalysisResults}} analyzes the results of the case studies and compares the outcomes of the DSS with the case study participants' shortlists of feasible DAO platforms. The results show that the DSS recommended nearly the same solutions as the case study participants suggested to their organizations after extensive analysis and discussions and does so more efficiently.

\end{itemize}

Section~\ref{Discussion} highlights barriers to the knowledge acquisition and decision-making process, such as motivational and cognitive biases, and argues how we have minimized these threats to the validity of the results. Section~\ref{Relatedwork} positions the proposed approach in this study among the other DAO platform selection techniques in the literature. Section~\ref{CONCLUSION} summarizes the proposed approach, defends its novelty, and offers directions for future studies.

\begin{figure}[!ht]
\centering
\caption{This figure is adopted from authors~\cite{Felix_Machart} and shows the layers of the blockchain stack. }
\includegraphics[trim=0 10 0 10,clip,width=0.42\textwidth]{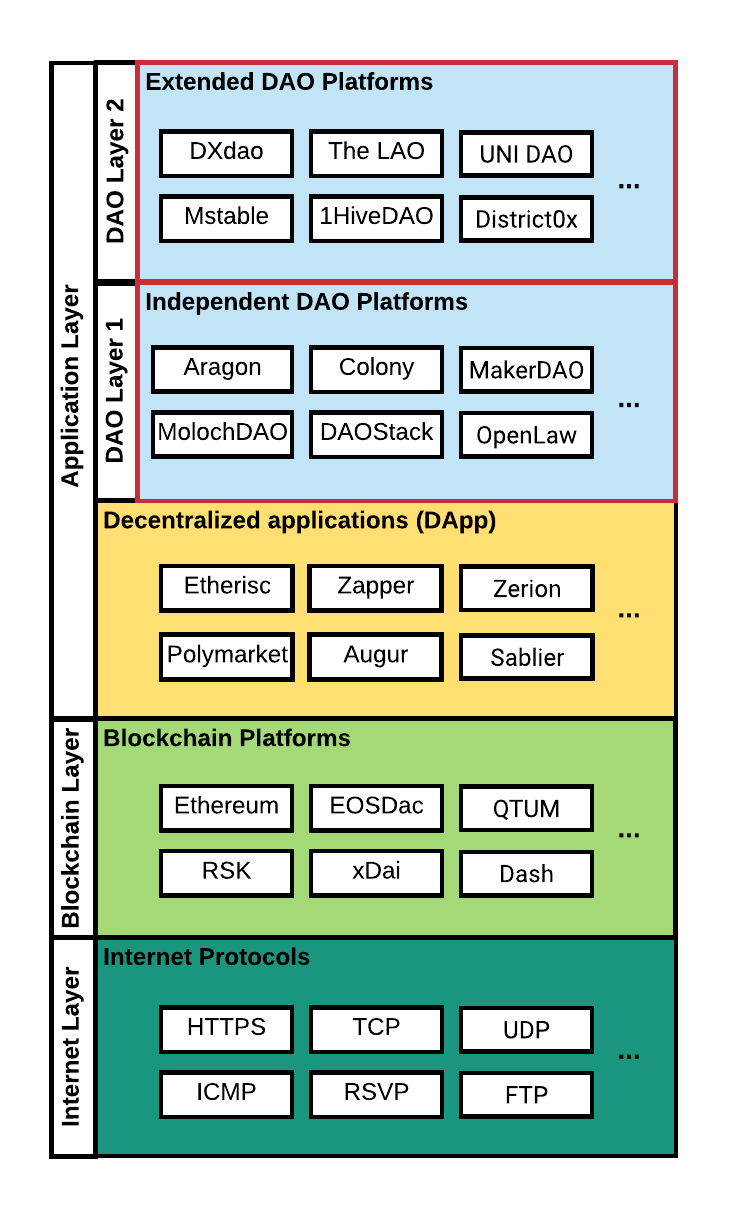}
\label{fig:DAOArchitecture}
\end{figure}

\section{Blockchain Technology Stack}\label{Background}
Decentralized applications based on blockchain technology operate within a larger ecosystem of Internet applications that run according to their protocols and rules. A blockchain technology stack~\cite{de2018governance} can be imagined to distinguish such applications and protocols. Each layer of the stack inherits the protocols and rules of the layer below, including the lower layers’ governance.  
Figure~\ref{fig:DAOArchitecture} illustrates the blockchain technology stack, which consists of the following main three main layers: the Internet, blockchain, and application layers. \newline

\subsection{The Internet layer}
Blockchain platforms~\cite{FarshidiBlockchain2019}, such as Bitcoin and Ethereum, exist at the bottom layer of the blockchain technology stack, but their operations depend on another technology stack: the Internet stack. Indeed, as a rule, blockchain platforms are unable to operate without Internet connectivity. Because these platforms operate on top of the Internet, their proper functioning is ultimately reliant on transmission control protocol/Internet protocol (TCP/IP)—the protocols responsible for routing and transferring packets between nodes on the Internet. Accordingly, decisions at the Internet level can have a dramatic impact on the operation and governance of blockchain platforms built on top of the Internet stack~\cite{Felix_Machart}. \newline

\subsection{The blockchain layer}
As aforementioned, blockchain platforms operate on top of the Internet layer and inherit the capabilities and limitations of that underlying layer, including its technical architecture and governance processes. While Internet Service Providers (ISPs) are responsible for routing packets through the Internet according to specific protocols (e.g., TCP/IP and border gateway protocol or BGP), nodes in a blockchain platform are responsible for validating and recording transactions into the underlying blockchain according to a particular set of rules. Each blockchain platform also introduces its own design decisions and governance mechanisms, including designing the underlying peer-to-peer network and the consensus protocol that facilitates agreement between the various nodes of the platform. For example, Bitcoin miners operate according to the Bitcoin proof-of-work protocol, which stipulates that miners should always add to the “longest chain” as defined by the amount of hashing power required to compute the chain \cite{de2018governance}.\newline

\subsection{The application layer} 
Generally speaking, an application layer is an abstraction level that masks the technical details of a communication channel and serves as a user interface on a network. The application layer in the blockchain technology stack focuses on developing blockchain solutions for use across different applications and industries. This layer contains three sub-layers: Decentralized applications (DApp), DAO Layer 1, and DAO Layer 2. 

\subsubsection {Decentralized applications}
In Ethereum, developers can develop their web applications, known as Decentralized Applications (DApps), without knowing the underlying mechanisms such as peer-to-peer networks, blockchain, and consensus rules. DApps are composed of at least: (1) a smart contract that acts as a software agent running on the blockchain performing predefined or preapproved tasks without any human involvement, and which is under the control of a set of business rules~\cite{swan2015blockchain}; and (2) a web user interface or web app allows users to interact with content through a web browser.

DApps in Ethereum may also use decentralized storage, such as the Inter-Planetary File System (IPFS), where data, files or websites, are stored in a distributed file system. DApps may also use a decentralized message service, such as Whisper, which provides one-to-one communication and acts as a 'decentralized chat' on the Ethereum platform, working on a peer-to-peer protocol, without servers but also without blockchains involved in the process~\cite{mohanty2018ethereum}.

From the Ethereum point of view, a DAO could be defined as a DApp that may be composed of other DApps, all running on the Ethereum platform, and whose business logic is encoded in terms of one or more smart contracts. In this way, the code (i.e., smart contracts) enforces at least part of a specific DAO's governance rules (i.e., decision-making rules). The fundamental distinction is the word "Autonomous". DAO could be viewed as one type of DApp, the fully autonomous DApp. Note, DApp is not necessarily DAO~\citeONLINE{dAppdifferent:online}.

\subsubsection {DAO Layer 1}
The development process of a DAO is significantly complicated, even for experienced software practitioners. One of the fundamental challenges for the Ethereum community is the lack of compliance standards and practical use cases of DAOs deployed on the blockchain, particularly when comparing their added value (e.g., efficiency or new services)to traditional and centralized organizations. As a response to this issue, some open-source software frameworks or independent DAO platforms have emerged to facilitate the implementation of DAOs~\cite{valiente2020evaluating}.

Independent DAO platforms represent toolkits to more easily spin up organizations on public blockchains that partly provide opinionated decision making and market mechanisms, partly leave everything open for the developer. Community and team members have been discussing that combining the frameworks could be promising as well, while some mechanisms pioneered by one framework team might also be implemented by another or a developer building a module on top.
These platforms that provide DAO deployment as-a-service allow users with insufficient knowledge on how blockchain works to create a DAO using a template that typically can be customized~\cite{el2020overview}.

Independent DAO Platforms are no-code platforms (such as Aragon, DAOstack, DAOhaus, and Colony) and provide tools to coordinate community resources allocation without the need for a central point of contact and a high degree of technical aptitude. For instance, Aragon is a software framework or Independent DAO Platform oriented to developing DAOs built on the Ethereum platform and creating configurable governance structures. Aragon states that it “gives internet communities unprecedented power to organize around shared values and resources” \citeONLINE{Nextleve54:online} and that it is software for creating and governing organizations such as clubs, companies, gaming guilds, cooperatives, nonprofits, open-source projects, and any other type of organization~\citeONLINE{AboutAra17:online}. DAOstack is another example of such independent DAO platforms. It is an open-source, modular DAO project, which leverages the technology and adoption of decentralized governance, enabling people to create the DApps(decentralized apps), DAOs, and DAO tools. ~\citeONLINE{DAOstack93:online}. As the last example, Colony is a DAO framework based on a reputation system (the user reputation weights, i.e., decision power). It aims to help organizations create their DAOs, named ‘colonies’, providing financial management, ownership, structure, and authority. The Colony network is composed of a suite of smart contracts which are deployed on the Ethereum blockchain \cite{valiente2020evaluating}.

\subsubsection{DAO Layer 2}
Extended DAO platforms have been developed based on DAO platforms of Layer 1. For instance, the dxDAO is built on top of DAOstack, and Reputation is the intrinsic DAOstack voting power signifier. District0x is a platform of decentralized markets that have been built based on Aragon. 

Initially, organizations can build their own DAO by using DAO platforms of Layer 1 and then extend it based on services offered by DAO platforms of Layer 2. For instance, Aragon-based DAOs can extend their functionality using pre-installed Aragon apps or modules as following~\cite{valienteresults}: \noindent\textit{Tokens} manage membership and voting power in a DAO, with the ability to mint (i.e., create) new tokens, assign existing tokens, and create vestings (i.e., tokens that are held aside for a while for the team, partners, advisors, and others who are contributing to the development of the project, and that can be released later). \noindent\textit{Voting mechanisms} create votes that execute actions on behalf of token holders, with the ability to see all open and closed votes, start a new vote and token poll holders in a DAO about a specific issue. \noindent\textit{Finance management modules} handle assets of a DAO, budget expenses, and record final transactions to have a history of past transfers, with the ability to create new transfers from this module. \noindent\textit{Agents} interact directly with any other smart contract on the Ethereum platform.

\section{Research Approach}\label{Research Approach}

Knowledge acquisition is the process of capturing, structuring, and organizing knowledge from multiple sources~\cite{gruber1989automated}. Human experts, discourse, internal meetings, case studies, literature studies, or other research methods are the primary sources of knowledge. The rest of this section outlines the research questions and elaborates on a mixed research method based on design science research, expert interviews, documentation analysis, and case study research to capture knowledge regarding DAO platforms, to answer the research questions, and to build a decision model for the DAO platform selection problem.

\begin{figure}
\centering
\caption{This figure shows an MCDM approach for the DAO platform selection problem in a 3-dimensional space. Note, the degree of the decision-makers' satisfaction with a solution according to their priorities and requirements (e.g. voting mechanism, decentralized type) ranges between the best and worst fit solutions (e.g. Aragon, DAOStack), which is represented by a range of colors from red to dark green.}
\includegraphics[trim=20 80 130 120,clip,width=0.45\textwidth]{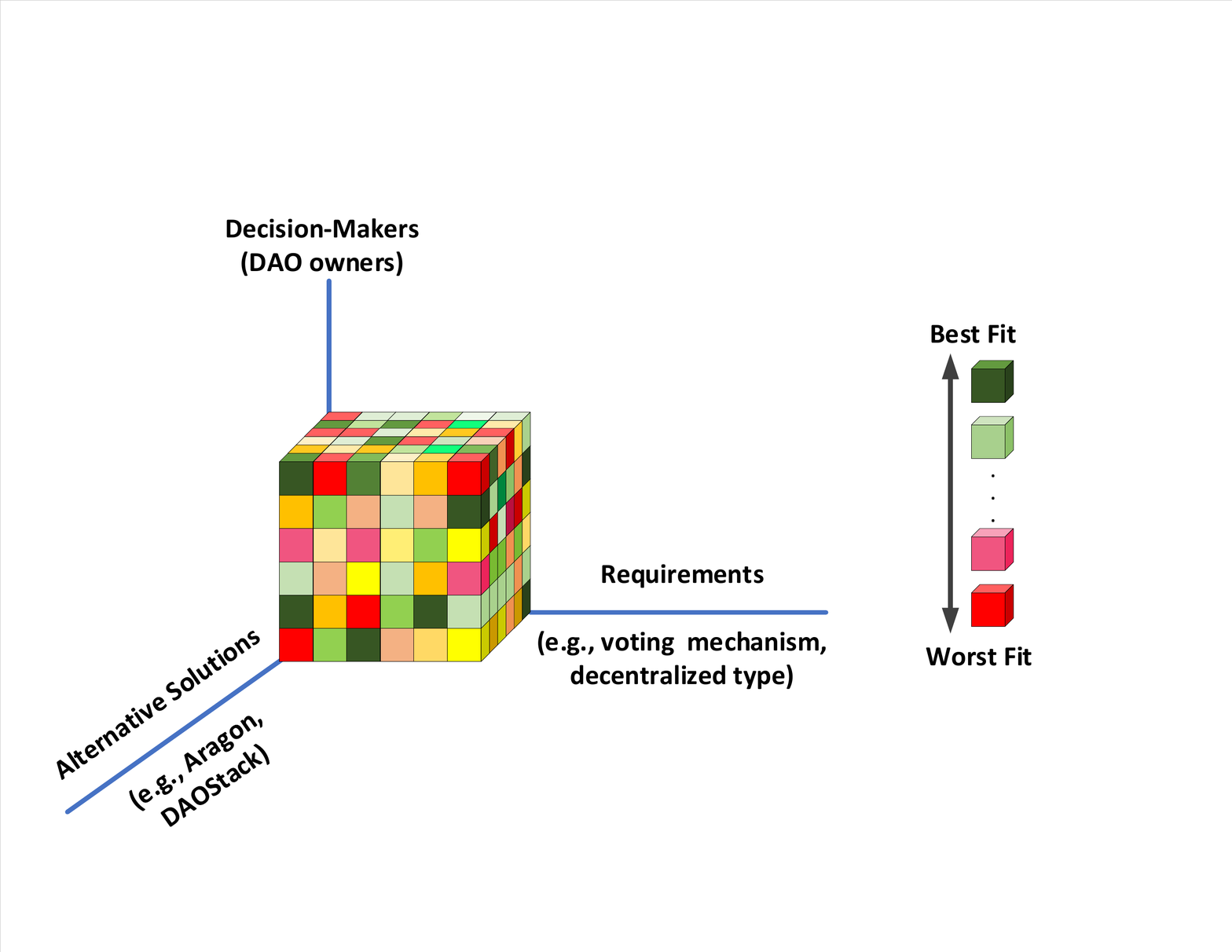}
\label{fig:3DSpace}
\end{figure}

\subsection{Problem Definition}

In this study, we formulate the DAO Platform selection problem as an MCDM problem:

Let $Platforms=\{p_1,p_2, \dots p_{|Platforms|}\}$ be a set of DAO platforms in the market (i.e., Aragon, DAOStack, and Colony). Furthermore, $Features=\{f_1,f_2, \dots t_{|Features|}\}$ be a set of DAO features (i.e., Decentralization Types, Voting Mechanism) of the DAO platforms, and each platform $p$, where $p \in Platforms$, supports a subset of the set $Features$. The goal is finding the best fitting DAO platforms as solutions, where $Solutions \subset Platforms$, that support a set of DAO feature requirements, called $Requirements$, where $Requirements \subseteq Features$. An MCDM approach for the selection problem receives $Platforms$ and their $Features$ as its input, then applies a weighting method to prioritize the $Features$ based on the decision-makers' preferences to define the $Requirements$, and finally employs a method of aggregation to rank the $Platforms$ and suggests $Solutions$. Accordingly, an MCDM approach for the DAO platform selection problem can be formulated as follows:

\begin{equation}
\begin{aligned}
MCDM: Platforms \times Features \times \\ Requirements \to Solutions
\end{aligned}
\end{equation}

Typically, a unique optimal solution for an MCDM problem, including DAO platform selection, does not exist, and it is essential to apply decision-makers' preferences to differentiate between solutions~\cite{farshidi2020multi,triantaphyllou2000multi}. Particular DAO platforms might fit into an organization; however, some might be better than others. It is tough to state which DAO platform is the best one, partially because we can not predict the future or know how the organizations would have evolved if a different DAO platform was selected. Moreover, we must note that such a technology selection process can never be completely objective because humans, as the main decision-makers, have to make decisions.
Figure~\ref{fig:3DSpace} visualises the MCDM approach for the DAO platform selection problem in a 3D space. It shows that the degree of satisfaction of the decision-makers with a suggested solution is fuzzy, which means that the satisfaction degree from a decision-maker perspective may range between completely true (best fit) and completely false (worst fit)~\cite{dvovrak2018affordance} which is represented by a range of colors from red to dark green.

\subsection{Research Questions}\label{DAO_ResearchQuestion}
The Main Research Question (MRQ) of this study is as follows:\newline

\noindent\textbf{MRQ:} How can knowledge regarding DAO platforms be captured and organized systematically to support decentralized autonomous organizations with the decision-making process?\newline

We formulated the following research questions to capture knowledge regarding the DAO platform systematically and to build a decision model for the decision problem based on the framework~\cite{farshidiCSP}:\newline

\begin{itemize}
\item[]\textbf{$RQ_1$:} Which DAO concepts should be considered as DAO features in the decision model?
\item[]\textbf{$RQ_2$:} Which DAO platforms should be considered in the decision model?
\item[]\textbf{$RQ_3$:} Which software quality attributes can be used to evaluate DAO platforms?
\item[]\textbf{$RQ_4$:} What are the impacts of DAO features on the quality attributes of DAO platforms?
\item[]\textbf{$RQ_5$:} Which DAO platforms currently support the DAO features?
\end{itemize}

\subsection{Research Method}

Research methods are classified based on their data collection techniques (interview, observation, literature, etc.), inference techniques (taxonomy, protocol analysis, statistics, etc.), research purpose (evaluation, exploration, description, etc.), units of analysis (individuals, groups, process, etc.), and so forth~\cite{meredith1989alternative}. Multiple research methods can be combined to achieve a fuller picture and a more in-depth understanding of the studied phenomenon by connecting complementary findings that conclude from the methods from the different methodological traditions of qualitative and quantitative investigation~\cite{johnson2004mixed}. 


Recently, we designed a framework~\cite{farshidiCSP} and implemented a DSS~\cite{farshidi2018DSS} for supporting software engineers (decision-makers) with their MCDM problems in software production. The framework provides a guideline for decision-makers to build decision models for MCDM problems in software production following the six-step of the decision-making process~\cite{Majumder2015}: (1) identifying the objective, (2) selection of the features, (3) selection of the alternatives, (4) selection of the weighing method, (5) applying the method of aggregation, and (6) decision making based on the aggregation results. In this study, we used the framework to build a decision model for the DAO platform selection problem. We employed design science, expert interviews, and document analysis as a mixed data collection method to capture DAO platforms' knowledge and answer the research questions.

\subsection{Design science} 
\textit{Design science} is an iterative process~\cite{simon2019sciences} has its roots in engineering~\cite{hevner2004design}, is broadly considered a problem-solving process~\cite{fortus2005design}, and tries to generate generalizable knowledge concerning design processes and design decisions. The design process is a collection of hypotheses that can ultimately be proven by developing the artifact it describes~\cite{walls1992building}. The research approach for building decision models for MCDM problems in software production is Design Science, which addresses research by developing and evaluating artifacts to meet defined business requirements~\cite{hevner2008design}.

In the previous study, we designed a theoretical framework and implemented a DSS for supporting software practitioners with their MCDM problems in software production~\cite{farshidi2020multi}. This study employs the framework to build a decision model for the DAO platform selection problem. Additionally, we employed the DSS to reduce the cost of the decision-making process. 

\subsection{Expert Interviews} 
\textit{Expert Interview} is an essential knowledge acquisition technique~\cite{chen2004electrical} in qualitative research. The main source of knowledge to build a decision model based on the framework~\cite{farshidiCSP} is domain expertise. A series of qualitative semi-structured interviews based on Myers' and Newman's guidelines~\cite{myers2007qualitative} has been conducted to explore tacit knowledge of domain experts regarding DAO Platforms and evaluate the outcomes of our study. Ten domain experts, including DAO developers, decentralized autonomous organizations, and blockchain experts from different organizations, have participated in the research to assist us with answering the research questions. Note that this set of interviews were different from the interviews we conducted during the case study research with the case study participants. 
 
A role description was developed before contacting potential domain experts to assist their expertise and to ensure the right target group. Next, we contacted the selected experts by email using the role description and information about our research topic. Note, the expert selection process has been done pragmatically and conveniently based on the reported expertise and experience mentioned on the \textit{LinkedIn} profile of the experts. We considered a set of expert evaluation criteria (including ``Years of experience'', ``Expertise'', ``Skills'', ``Education'', and ``Level of expertise'') to select the experts.  

Each expert interview followed a semi-structured interview protocol (See Appendix~\ref{InterviewProtocols}) and lasted between 45 and 60 minutes. Additionally, we used a number of open questions to elicit as much information as possible from the experts minimizing prior bias. All interviews were done virtually through meeting platforms, such as Skype and Zoom, and recorded with the interviewees' permission, then transcribed for further analysis. 

Captured knowledge after each interview was typically propagated to the next one to validate the acquired knowledge incrementally. Finally, our findings and interpretations were sent back to the interview participants for their final approval. Note, for the validity of the results, the research's data collection phases were not affected by the case study participants; furthermore, none of the interviewees or researchers were involved in the case studies. 

\subsection{Document analysis} 

\textit{Document analysis} is a systematic procedure for reviewing or evaluating documents, including manuscripts and illustrations, that have been published without a researcher's intervention~\cite{bowen2009document}. Document analysis is one of the analytical methods in qualitative research that requires data investigation and interpretation to elicit meaning, gain understanding, and develop empirical knowledge~\cite{corbin2014basics}. 

There is not a significant number of academic literature available about DAO platforms and related concepts due to their novelty and, in part, due to the fast growth and development of the industry. For this reason, we have also added gray literature to our knowledge base, which resulted in a significant increase in the amount of information we could find. Currently, around 59\% of the sources are web pages, 11\% are peer-reviewed articles, and 16\% are documentation of the platforms themselves. The rest (14\%) are a collection of videos, white papers, forum discussions, and books. In these sources, we specifically identified features from each of the platforms.

\begin{figure}[!ht]
  \centering
    \caption{This figure shows the distribution of the selected studies in the document analysis phase based on their publication year.}
    \includegraphics[trim=10 40 25 40,clip,width=0.45\textwidth]{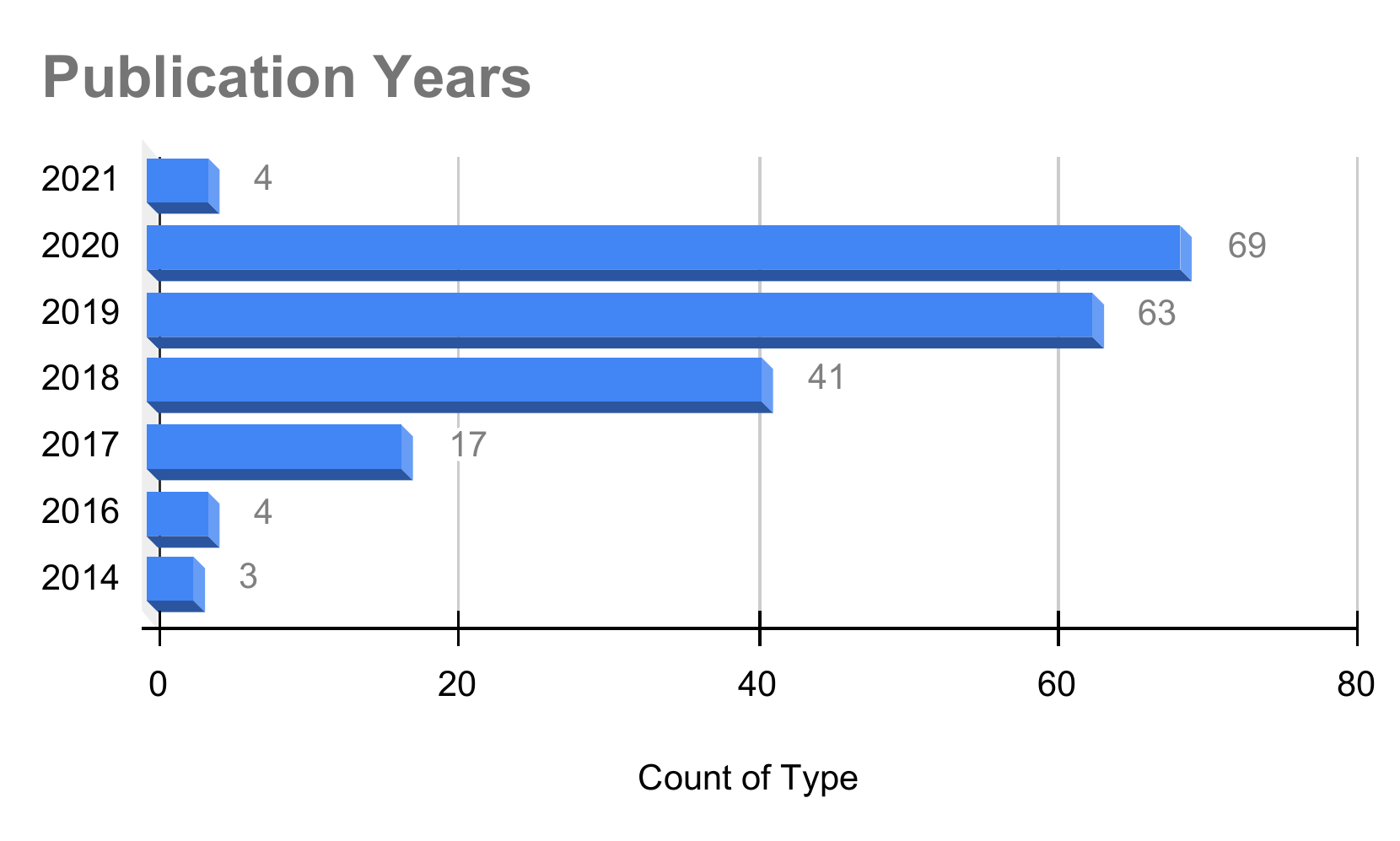}
  \label{fig:PublicationYears}
\end{figure}

It is essential to highlight that the selected sources of knowledge in the document analysis phase of this research that discuss the DAO platforms are spread across the early years of the emergence of the DAO concepts (2014)~\cite{buterin2014next} to the present (2021). Additionally, the possibility of existing trends among software practitioners and researchers in selecting DAO platforms has been investigated in this study. Accordingly, we observed that DAO platforms and their selection process gained more attention in the past four years (see Figure~\ref{fig:PublicationYears}). 

We created an extraction form for collecting knowledge and ensure that they are consistent with relevant knowledge, and also we checked that the knowledge gathered answered the research questions. The collected knowledge, which corresponds to the research questions, have been classified into five categories:  \textit{DAO platforms}, \textit{DAO features}, \textit{mapping among the DAO features and the quality attributes}, \textit{Quality Attributes} and \textit{mapping among the DAO features and the DAO platforms}. 

\subsection{Case Study}
\textit{Case study research}~\cite{jansen2009applied} is an empirical methodology that investigates a phenomenon within a particular context in the domain of interest. Case study research can be employed for collecting data regarding a particular phenomenon or for applying and evaluating a tool to understand its efficiency and effectiveness using interviews. Yin~\cite{yin2017case} identifies four types of case study designs based on holistic versus embedded and single versus multiple. This study employs holistic multiple case designs: examining multiple real-world decentralized autonomous organizations as multiple cases within their context to understand one specific unit of analysis and evaluating the decision model for the DAO platform selection problem.

\noindent\textbf{Objective:} Building a valid decision model for the DAO selection problem was the main goal of this research.\newline
\noindent\textbf{The cases:} The analysis units were three industry case studies, performed in the Netherlands, the United States, and Iran, in the context of three decentralized autonomous organizations.\newline
\noindent\textbf{Methods:} We conducted multiple expert interviews with the case study participants to understand their requirements, concerns, and preferences regarding the DAO platform selection problem.\newline
\noindent\textbf{Selection strategy:} In this study, we selected \textit{multiple case study}~\cite{yin1981case} to analyze the data both within each situation and across situations, to more extensive exploring the research questions and theoretical evolution, and to create a more convincing theory.\newline
\noindent\textbf{Theory:} The proposed decision model is a valid reference model to support decentralized autonomous organizations with the DAO platform selection problem.\newline
\noindent\textbf{Protocol:} To conduct the case studies and evaluate the proposed decision model, we followed the following protocol:
\begin{itemize}
  \item[] \noindent\textbf{Step 1. Requirements elicitation: } The participants defined their DAO language feature requirements and prioritized them based on the \textit{MoSCoW} prioritization technique~\cite{atern2008handbook}. Furthermore, they identified a set of DAO platforms as potential solutions for their DAOs.
  \item[] \noindent\textbf{Step 2. Results and recommendations: } We defined three separate cases in the knowledge base of the DSS according to the case studies' requirements and priorities. Next, the DSS recommended a set of feasible DAO platforms as alternative solutions per case individually. Next, the outcomes were discussed with the case study participants.
  \item[] \noindent\textbf{Step 3. Analysis: } We compared the DSS suggested feasible solutions with the case study participants' preselected solutions that they had suggested to their organizations based on extensive analysis. Furthermore, we analyzed the outcomes and observations and then reported them to the case study participants and subsequently received their feedback on the results.
\end{itemize}

\section{Multi-Criteria Decision-Making} \label{MCDM}

\begin{figure*}[!ht]
  \caption{This figure is adapted from this study~\cite{farshidiCSP} and shows the main building blocks of the decision support system beside the proposed decision model for the DAO Platform selection problem.}
  \centering
   \includegraphics[trim=10 305 280 15,clip,width=1.0\textwidth]{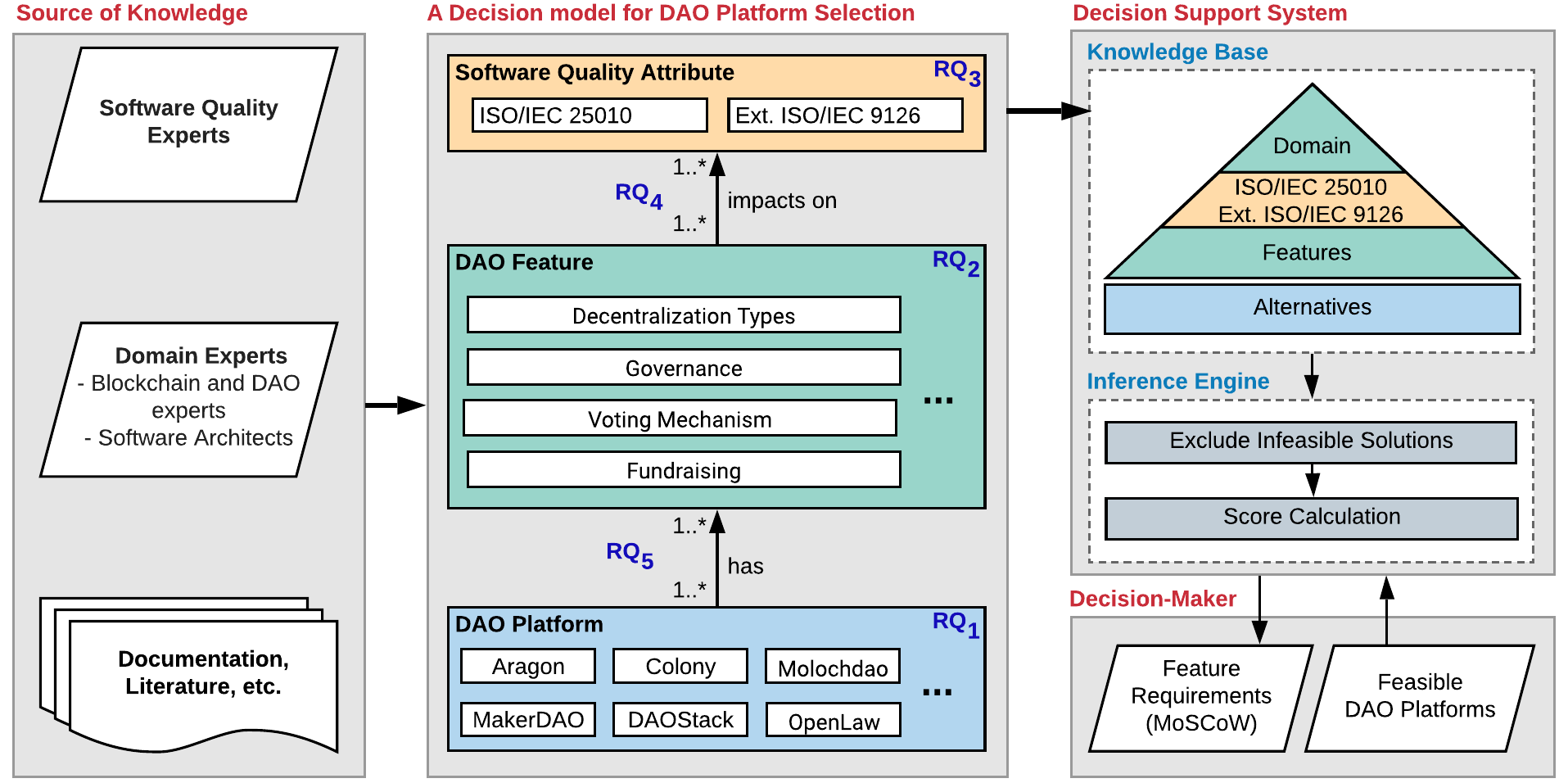}
  \label{fig:MainModel}
\end{figure*}

We follow the framework~\cite{farshidi2020model} as modeled in Figure~\ref{fig:MainModel} to build a decision model for the DAO platform selection problem. Generally speaking, a decision model for an MCDM problem contains decision criteria, alternatives, and mappings. Figure~\ref{fig:MainModel} represents the main building blocks of the decision support system besides the proposed decision model.\newline

\subsection{\textit{RQ1: DAO Platforms}}

\begin{table}[!ht]
  \caption{This figure shows the DAO platforms that were mentioned at least five sources of knowledge (communities and domain experts). This list has been considered as the DAO platform alternatives in the decision model.}
  \centering
   \includegraphics[trim=80 250 70 50,clip,width=0.48\textwidth]{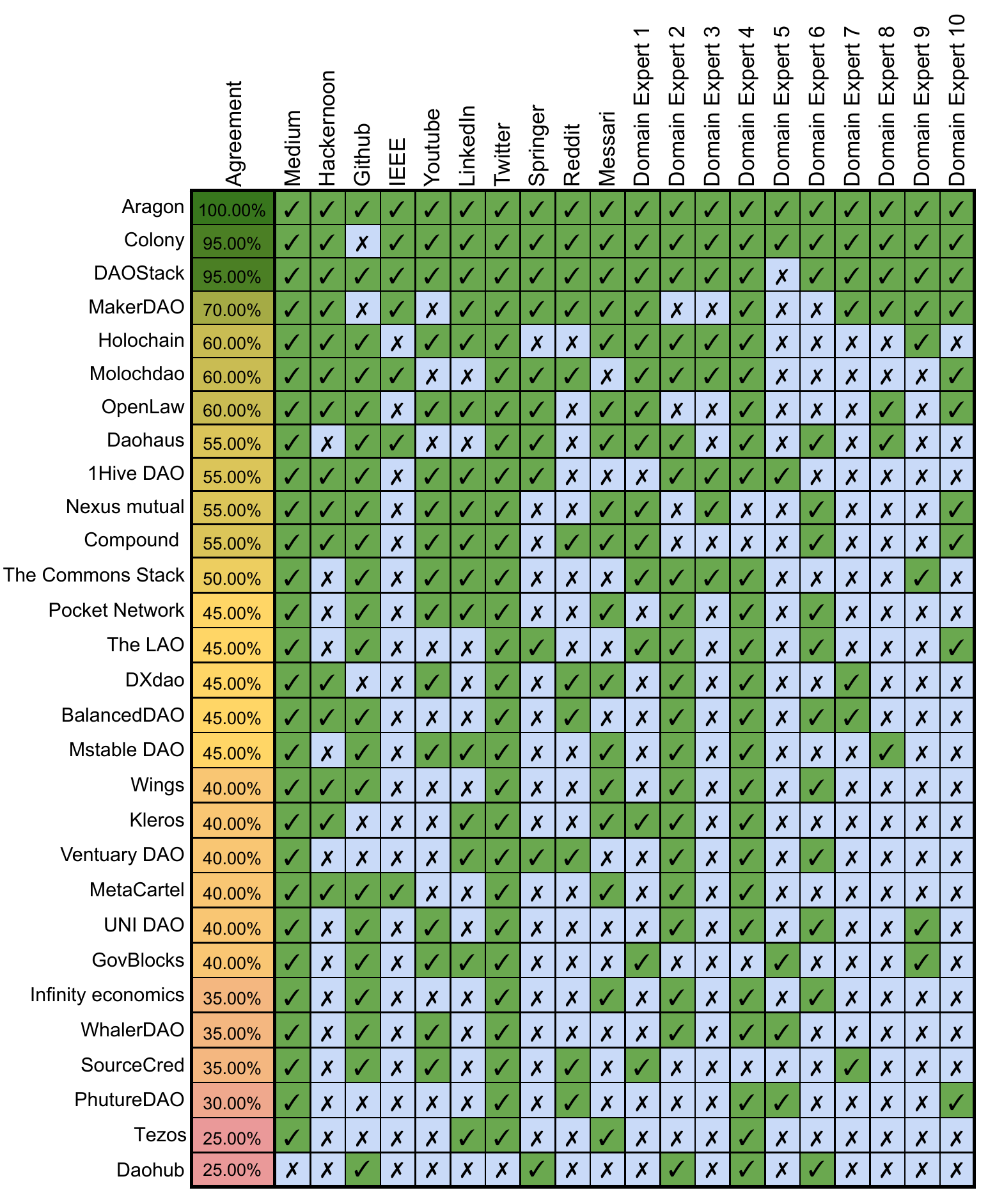}
  \label{tbl:Alternatives}
\end{table}

We identified 90 DAO platforms as our initial hypothesis and ended up with 28 DAO platforms based on literature and expert interviews to answer the first research question. Accordingly, we explored literature based on the following search keywords: ``DAO'', ``DeFi'', ``DAO As Service'', and ``decentralized autonomous organization'' platforms. Additionally, we exploit our network of domain experts, including software engineers and academics.  We reviewed the published surveys and reports from well-known communities, including Medium~\citeONLINE{Medium:online}, Github~\citeONLINE{Github:online}, IEEE~\citeONLINE{IEEE:online}, Hackernoon~\citeONLINE{Hackernoon:online}, YouTube~\citeONLINE{Youtube:online}, LinkedIn~\citeONLINE{LinkedIn:online}, Twitter~\citeONLINE{Twitter:online}, Springer~\citeONLINE{Springer:online}, Reddit~\citeONLINE{Reddit:online}, Messari~\citeONLINE{Messari:online}.

We conducted a set of expert interviews with ten experts to gain more insight into the popular and applicable DAO platforms and to evaluate our findings. It is interesting to highlight that most of the domain experts were familiar with a limited number of the list's DAO platforms (See the ten Domain Experts' columns on Table~\ref{tbl:Alternatives}). We only considered the DAO platforms mentioned on at least five sources of knowledge (including communities and domain experts) to prevent potential biases. Finally, we analyzed the data and ended with 28 alternative DAO platforms mentioned on at least three resources. Table~\ref{tbl:Alternatives} shows the complete list of the DAO platforms that we have selected in the decision model. 

\subsection{\textit{RQ2: DAO Features}} 
Domain experts were the primary source of knowledge to identify the right set of DAO features, even though documentation and literature studies of DAO platforms can be employed to develop an initial hypothesis about the DAO feature set. Each DAO feature has a data type, such as \textit{Boolean} and \textit{non-Boolean}. For example, the data types of DAO features, such as the \textit{popularity in the market} and supportability of \textit{Quadratic voting}, can be considered as \textit{non-Boolean} and \textit{Boolean}, respectively. 

The initial set of DAO features was extracted from the following sources: web pages, white papers, scientific papers, documentation, forum discussion, books, videos, and dissertation. we distinguished 118 \textit{Boolean} and ten \textit{non-Boolean} features in our initial list. Afterward, ten domain experts have participated in this phase of the research to identify a potential list of DAO features.

Accordingly, 77 Boolean and five non-Boolean DAO features\footnote{The entire lists of the DAO features and their mapping with the considered DAO platforms are available and accessible on the \textit{DAO Platform Selection} website (\url{https://dss-mcdm.com})} were identified and extracted from the  outcomes of the expert interviews. Eventually, the validity and reliability of the final list of the DAO features were evaluated and confirmed by the domain experts.

\subsection{\textit{RQ3: Software Quality Attributes}}
According to IEEE Standard Glossary of Software Engineering Terminology~\cite{IEEEStandardQA,samadhiya2010quality}, the quality of software products is a model to which a system, component, or process meets defined requirements (such as availability, scalability, security, and Operability) and the degree to which a system, component or process addresses the requirements or expectations of a user. It is essential to find quality attributes widely supported by other researchers to estimate the system's characteristics.

The ISO/IEC 25010~\cite{iso2011iec25010} presents best practice recommendations based on a quality assessment model. The quality model defines which quality characteristics should be considered when assessing a software product's properties. 
A set of quality attributes should be specified in the decision model~\cite{farshidiCSP}. In this study, we used the \textit{ISO/IEC 25010} standard~\cite{iso2011iec25010} and \textit{extended ISO/IEC 9126} standard~\cite{carvallo2006extending} as two domain-independent quality models to investigate DAO features based on their impact on quality attributes of DAO platforms. The key rationale behind employing these software quality models is that they are a standardized way of measuring a software product. Furthermore, they explain how efficiently and reliably a software product can be used. 

The literature study results show that researchers and practitioners do not agree upon standard criteria for assessing DAO platforms, including quality attributes and features. See Table~\ref{DAO_tableLiterature}).

The last columns of Table~\ref{DAO_tableLiterature} denote the outcomes of the analysis concerning the common criteria and alternatives of this study with the selected studies. Let us define the coverage of the i-th selected study as follows:

$Coverage_i= \dfrac{CQ_i+CF_i}{C_i}\times 100$

Where $CQ_i$ and $CF_i$ signify the numbers of common quality attributes (column \#CQ) and features (column \#CF) of the i-th selected study, respectively, furthermore, $C_i$ defines the number of suggested criteria by the i-th selected study. The last column (\textit{Cov.}) of Table~\ref{DAO_tableLiterature} shows the percentage of the coverage of the considered criteria within the selected studies. On average, 79.24\% of those criteria are already considered in this study.

\subsection{\textit{RQ4: Impacts of the DAO Platform Features on the Software Quality Attributes }}
The mapping between the sets \textit{software quality attributes} and \textit{DAO platforms} was identified based on domain experts' knowledge. Four domain experts participated in this phase of the research to map the DAO platforms ($Features$) to the quality attributes ($Qualities$) based on a Boolean adjacency matrix ($ Qualities \times Features \to Boolean$). For instance, \textit{Infrastructure decentralization} as a DAO feature influences the \textit{Operability} quality attribute. The framework does not enforce a DAO feature in a single quality attribute; DAO features can be part of many quality attributes. 
\begin{itemize}
\item \textbf{Functional appropriateness} is defined in ISO/IEC 25010 as the degree to which the functions of DAO platforms facilitate the accomplishment of specified tasks and objectives.

\item \textbf{Operability} is the degree to which a DAO platform has attributes that make it easy to operate and control.

\item \textbf{Interoperability} define the degree to which two or more DAO platforms can exchange information and use the information that has been exchanged.

\item \textbf{Functional correctness} defines the degree to which a DAO platform provides the correct results with the needed degree of precision.

\item \textbf{Ownership} Attributes  in  relation  to  the  intellectual  property rights. 
\item \textbf{Functional completeness} is the degree to which the set of functions of DAO platforms covers all the specified tasks and user objectives.
\end{itemize}

\subsection{\textit{RQ5}: Supportability of the DAO Platform Features by the DAO Platforms}
A DAO platform contains a set of DAO features that can be either Boolean or non-Boolean. A Boolean DAO feature ($Feature^{B}$) is a feature that is supported by the DAO platform, for example, supporting the \textit{Quadratic Voting}. A non-Boolean DAO feature ($Feature^{N}$) assigns a non-Boolean value to a particular DAO platform; for example, the popularity in the market of a DAO platform can be ``high'', ``medium'', or ``low''. Accordingly, this study's DAO features are a collection of Boolean and non-Boolean features, where $ Features= Feature^{B} \cup Feature^{N}$.

The mapping $BFP: Feature^{B} \times Platforms\to \{0,1\}$ defines the supportability of the Boolean DAO features by the platforms. So that $BFP(f,p)=0$ means that the platform $p$ does not support the DAO feature $f$ or we did not find any evidence for proof of this feature's supportability by the DAO platform. Moreover, $BFP(f, l)=1$ signifies that the platform supports the feature. The mapping \textit{BFP} is defined based on documentation of the  DAO platforms and expert interviews.  Tables~\ref{fig:BFA1} and~\ref{fig:BFA2} in the appendix present the Boolean Features that we have considered in the decision model.

The experts defined five non-Boolean DAO features, including ``Popularity in the market'', ``Maturity level of the company'',  ``Developer Resources (People)'', ``Upgradability'',and ``Scalability''. The assigned values to the non-Boolean DAO features for a specific DAO platform is based on a 3-point Likert scale (\textbf{H}igh, \textbf{M}edium, and \textbf{L}ow), where $ NFP: Features^{N} \times Platforms\to \{H, M, L\}$, based on several predefined parameters. For instance, the ``popularity in the market'' of a DAO platform was defined based on the following parameters: the number of the ``Google hits'', ``Twitter (follower)'', ``LinkedIn (follower)'', and the popular forums and reports that considered the platform in their evaluation. Table~\ref{fig:NFP_DAO} in the appendix shows the non-Boolean DAO features, their parameters, and sources of knowledge. 

\begin{table*}
 \caption{ This table shows the DAO feature requirements, based on the MoSCoW prioritization technique (Must-Have (M), Should-Have (S), Could-Have (C), and Won't-Have(W)). Note, the \textit{Coverage} column denotes the percentage of DAO platforms that support each feature. The most vulnerable features changed from \textit Must-Have to \textit Should-Have or \textit Won't-Have to None (Note ``$>$S'' indicates  that Must-have changed to Should-have and ``$>$N'' denotes Won't-Have to None.)}
 
  \label{tbl:MoSCoW}
  \centering
   \includegraphics[trim=0 0 0 0,clip,width=0.9\textwidth]{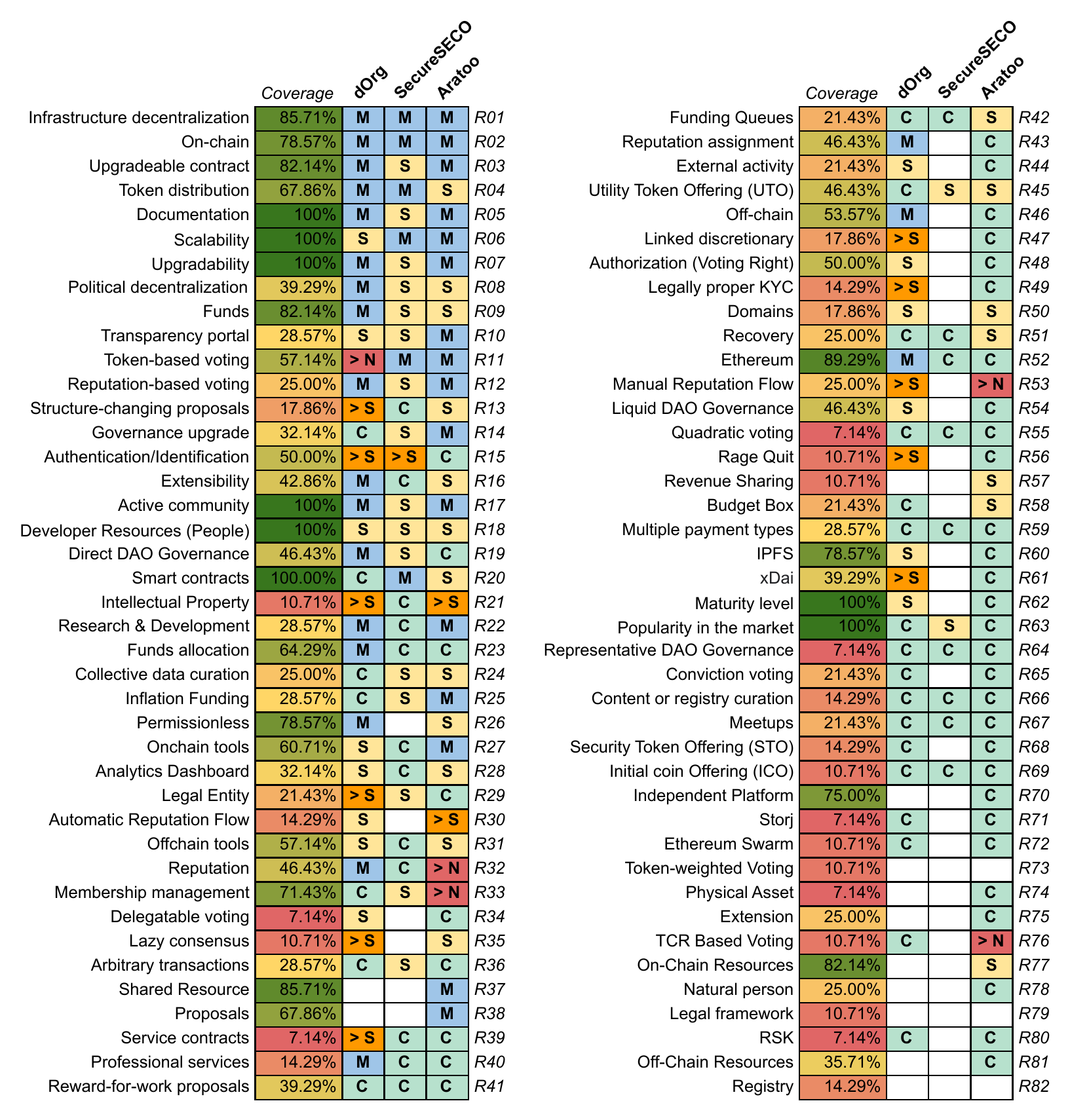}
\end{table*}

\subsection{DAO Feature Requirements} 
DSS~\cite{farshidi2018DSS} receives the DAO feature requirements based on the \textit{MoSCoW} prioritization technique~\cite{atern2008handbook}. 
Decision-makers should prioritize their DAO feature requirements using a set of weights ($W_{MoSCoW}= \{w_{Must}, w_{Should}, w_{Could}, w_{Won't}\}$) according to the definition of the \textit{MoSCoW} prioritization technique. DAO feature requirements with \textit{Must-Have} or \textit{Won't-Have} priorities act as hard constraints and DAO feature requirements with \textit{Should-Have} and \textit{Could-Have} priorities act as soft constraints. So that the DSS excludes all infeasible DAO platforms which do not support DAO features with \textit{Must-Have} and support DAO features with \textit{Won't-Have} priorities. Then, it assigns non-negative scores to feasible DAO platforms according to the number of DAO features with \textit{Should-Have} and \textit{Could-Have} prioritizes~\cite{farshidiCSP}.

\section{Empirical Evidence: The Case Studies}\label{CaseStudies}
Three industry case studies have been conducted to evaluate and signify the decision model's efficiency and effectiveness to address the DAO platform selection problem. We selected the case study organizations from three different domains, including web3 development, open-source software security, and decentralized finance (DeFi), for increasing variety in our evaluation. Moreover, the selected case study organizations were located in three different countries: the United States, Iran, and the Netherlands.

\subsection{Case Study 1: dOrg} 
dOrg is a decentralized autonomous collective of developers specialized in Web3 design and development that works with industry-leading projects. dOrg is a freelancer collective with over 25 members that work remotely to build solutions for dApps, smart contracts, prototypes, and experiences. The collective is run in a decentralized manner by worldwide DAO builders through the use of smart contracts. In other words, dOrg is a cooperative of blockchain software engineers that build DAO-related software. dOrg provides the DAO with continuous product development and operational support services. 

The experts at dOrg have created a Blockchain-Based Limited Liability Company (BBLLC): dOrg LLC, an Ethereum-based DAO with legal status in the United States. Doing that dOrg team demonstrated how a DAO could have official legal status, making it possible for DAOs to enter contractual agreements and offer participants liability protections.

The collective is releasing a new integration model for smart contracts that brings the simplicity of Web2 APIs to Web3. Web3 API enables app developers to interact with smart contract protocols from any programming language by GraphQL. It eliminates the need to embed language-specific Software Development Kits (SDKs) into dApps. dOrg has found this path to seamless app integration from numerous experiences bridging the gap between smart contracts and applications.

Their rigorous freelancer activation process and transparent workflow have resulted in collaborations with crypto brands such as Gnosis, eToro, Balancer, DAOstack, The Graph, and more. No centralized governing has the authority to control the interaction between freelancers and employers. There is direct contact between employers and employees without the need for any middleman. Every freelancer would be able to compete for any consignment which is posted on the platform. The cryptocurrency helps in eliminating the frauds involving a repudiation of payment and false claims on assets. The cryptocurrency wallets are secured, and payments can only be made by the wallet owner and do not need any information to be shared with any intermediary. The payments are made between crypto wallets which are highly secured by cryptographic keys. The network can be trusted because every transaction is verified by the network and subsequently written on an immutable ledger. The smart contracts release payments instantly once the transaction conditions are satisfied and there are no intermediate holders to delay payments. Anyone can apply for a job from anywhere in the world; there are no limitations on global access and payments. The smart contract holds the security amount and is trusted by both parties involved in the system as it is locked based on mathematical logic. Everyone in the network has access to a distributed ledger to verify that the system is working correctly at any instant. The experts at dOrg stated that they use DAOStack as their selected DAO platform to develop DAOs for their clients.  

\subsubsection{Requirements}

The experts at dOrg defined the following subset of requirements for their DAO (for more detail, see Table~\ref{tbl:MoSCoW}): 
\begin{itemize}
\item The dOrg needs the DAO that people decide on policy initiatives directly (R03).
\item The DAO must be able to use on-chain resources to allow a DAO to directly exert control and initiate action via a smart contract (R14, R15).
\item The DAO must support those token holders who can become contractors by submitting proposals for their project's funding by using the DAO funds (R18).
\item They need to approve the proposal in Research \& Development, Service contracts, Professional services, Reputation, Structure-changing proposals in their DAO (R28, R29, R30, R32, R36).
 \item \textit{Upgradability}, \textit{scalability}, and \textit{Maturity level} were part of their quality concerns, so that they preferred to hire highly mature DAO (R78, R79, and R80).
 \item Besides their current knowledge and experience, the developers' availability is an essential factor that has profoundly impacted their decision-making process (R82). 
\end{itemize}

\subsubsection{Results}
The case study participants identified 69 DAO feature requirements, including 17\% hard-constraint features (Must-Have) and 83\% soft-constraint features (Should-Have and Could-Have).  
Table~\ref{tbl:tableSolutions} shows that Colony was the first, Aragon was the second, and DAOStack was the third feasible platform for this case study.   

The DSS scored DAOStack as the third solution, and only it supports ``delegable voting '' as a should-have feature. Moreover, ``Automatic Reputation Flow'' is another should-have feature supported by Aragon and DAOStack, respectively, the second and the third solution.
because the effect of Could Have features 

 The Should-Have features have higher priorities than the Could-Have features, so DAO platforms that support more Should-Have features score higher. However, because the Could-Have features supported by Colony are a lot, the DSS offers Colony as the first solution.
\subsection{Case Study 2: SecureSECO} 
The cybersecurity project SecureSECO intends to make the worldwide software ecosystem a safer place by maintaining a distributed ledger of facts about software used in the field. The data collected and maintained in the ledger can be used to prevent vulnerabilities in a software configuration from being abused by malicious attackers. 

The SecureSECO project is a collaboration between five companies and five universities with over ten researchers and tool developers who collaboratively contribute to the vision of a safer and more secure worldwide software ecosystem. They perform academic research, participate in hackathons, and provide academic research data for other research groups about software.

The SecureSECO project collectively calculates trust metrics for software packages and software-producing organizations. SecureSECO follows the open-source mantra that if there are enough participants, any software problem can be solved, and it is the community that should be in charge of a trust mechanism. For this reason, decisions about the trust calculation mechanism are taken by the collective instead of a centralized entity. For this reason, SecureSECO is managed in a DAO. In this way, they can provide meta-data on software trustworthiness as a commons, similar to the air we breathe and the water we swim in.

\subsubsection{Requirements}
The experts at SecureSECO defined the following subset of requirements for Implementing the DAO (for more detail, see Table~\ref{tbl:tableSolutions}): 

\begin{itemize}
\item They need a DAO that should endure if some parts (computers, nodes, etc.) are broken (R01).

\item The DAO must provide a mechanism that people can buy tokens, vote, and sell the tokens (R11).

\item The DAO must support smart contracts that a set of roles are predefined by computer code in a smart contract, which is replicated and executed by all network nodes, and it should be upgradeable (R77, R03, R20).

\item They need a DAO that can configure and update its governance system (R14).

\item The popularity in the market (R63), Scalability (R06), Maturity level (R62), and upgradeability (R55) are the main quality concerns of the experts at SecureSECO when they want to select potential DAO.

\item The potential DAO should be mature enough and trendy in the market because they have comprehensive documentation and friendly communities (R05, R17).

\end{itemize} 

\subsubsection{Results}
The case study participants at SecureSECO identified 46 DAO feature requirements. They prioritized over 50\% of them as soft constraints (Could-Have and Should-Have) features based on their assumptions. The SecureSECO experts indicated \textit{Aragon}, \textit{Colony},  as their top potential DAO platforms.

Table \ref{tbl:tableSolutions} shows that Colony was the first feasible platform for SecureSECO. Additionally, Aragon, DAOStack, MakerDAO, Molochdao, and Kleros were scored as the second to sixth potential solutions.
They only identified a small number of Must-Have features and defined a limited number of Must-Have features. They identified Could-Have features more than others. Thus, the DSS had to suggest more feasible solutions.


\begin{table*}
 \caption{This table indicates the context of the case study companies (Context), the feature requirements (Requirements), the case study participants' ranked shortlists (CP). Moreover, the numbers of feature requirements (\#Feature Req) and the percentages of the MoSCoW priorities are shown in the table. For instance, the percentage of the Must-Have priority for \textit{dOrg} is 17\%, and finally, the outcomes of the DSS for the case studies (DSS Solutions), which are based on their requirements and priorities. These numbers in percentages in this section of the table signify the calculated scores by the DSS. For instance, the score of the \textit{Aragon} platform for \textit{SecureSECO} is 88\%.}
 
  \label{tbl:tableSolutions}
  \centering
   \includegraphics[trim=117 190 120 50,clip,width=0.9                                    \textwidth]{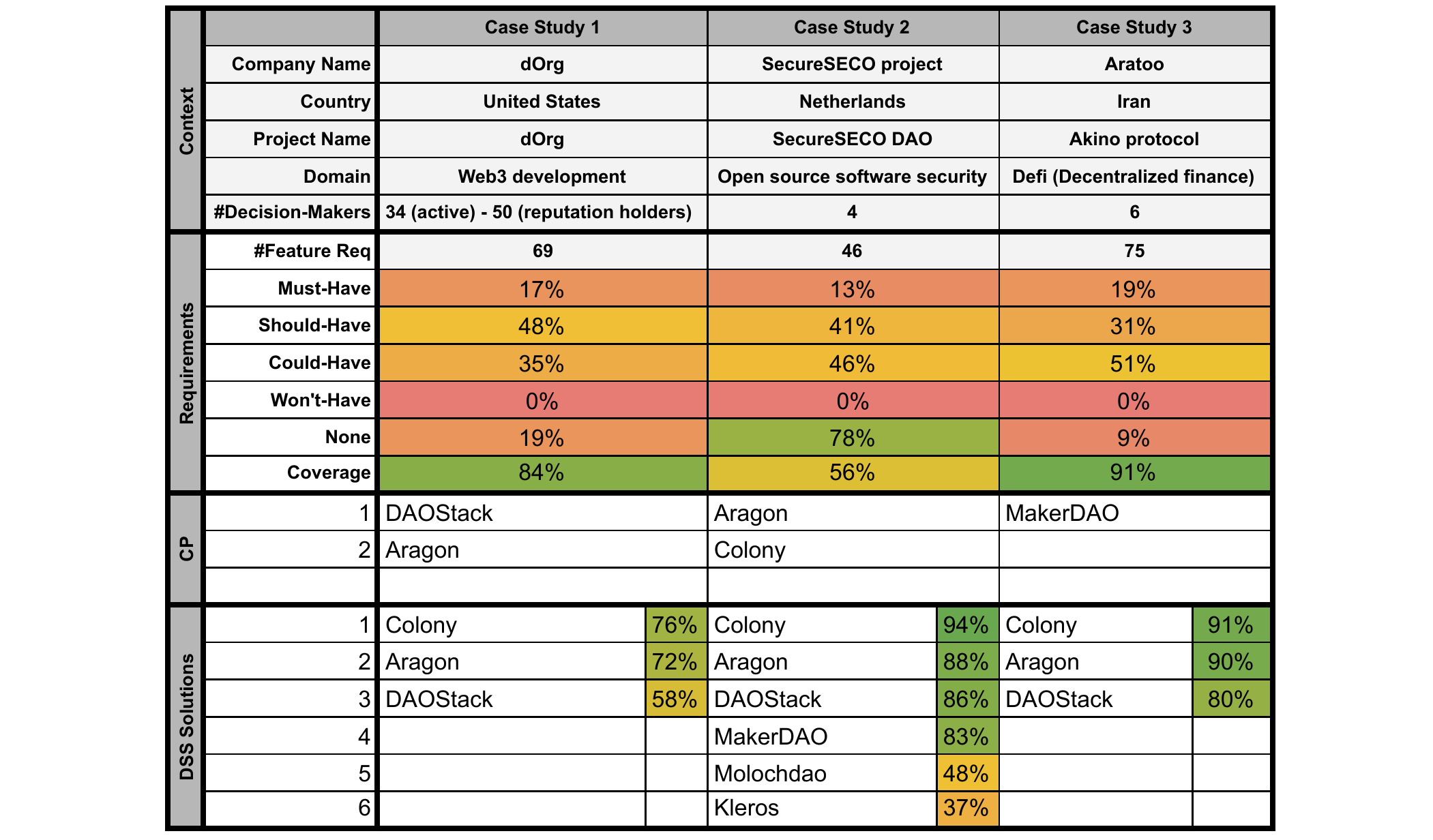}
\end{table*}

\subsection{Case Study 3: Aratoo} 
Aratoo is an Iranian decentralized autonomous organization that aims to liberate the Iranian cryptocurrency market. They are developing, amongst other products, a DeFi wallet for managing cryptocurrencies. A DeFi crypto wallet is a non-custodial wallet where the users have complete access and control of their private keys and funds. DeFi wallets are at the core of the concept ``be your own bank''.

Aratoo is a transparent platform that uses smart contracts and native protocol to reduce investment risk, increase profits, and expand blockchain technology and decentralized systems. They have employed DeFi ecosystems, such as MakerDAO and CurveDAO, for lending, borrowing, exchanging, and governing.

The DAO will allow liquidity providers to make decisions on adding new pools, changing pool parameters, adding token incentives, and many other aspects of the protocol. A pool is a smart contract that implements the StableSwap invariant and thereby allows for exchanging two or more tokens.

A DeFi wallet serves the primary purpose of allowing users to store their funds without reliance on a third party.

 The DAO is essential for the Aratoo because people can trust the governing to update feeds promptly and under changes that are made across the DeFi ecosystem accurately. In other words, the role of DAO is for protocol governance and value accrual.
There needs to be strong incentives for the people involved in the DAO to report accurate updates, vote on them, and maintain that reporting/voting behavior into perpetuity.

Aratoo needs a DAO for governance and control of the protocol admin functionality and implementation of the Voting App.

\subsubsection{Requirements}

The experts at this company indicated the following subset of requirements of their DAO (for more detail, see Table~\ref{tbl:MoSCoW}):

\begin{itemize}

\item They need a DAO that a single individual or organization does not control (R08).
\item They need a mechanism that investigates the grid resources' trustworthiness through a reputation system and then decides the results (R12).
\item The DAO must support that proposers receive an automatic reputation reward if their proposal passes (R30).
\item The DAO must provide a feature that an organization can manage its collective databases of objects and maintain their curation (R24).
\item The DAO must provide a feature that can set a rate at which a DAOs token is minted and a ceiling to the supply (R25).
\item They need an analytic dashboard that shows real-time system feedback (R28).
\item Scalability and upgradability of the DAO were two key quality concerns of the case study participants (R06, R07).
\end{itemize}

\subsubsection{Results}
The case study participants at Aratoo identified 75 DAO feature requirements, including 19\% hard-constraint features (Must-Have) and 82\% soft-constraint features (Should-Have and Could-Have).

The case study participants looked for a platform supporting ``Token distribution'' (R04) and ``Lazy consensus'' (R35) as two Should-Have features. Based on our assessment, Aragon, Colony, and DAOStack support both of these features. ``Revenue Sharing'' (R57) as a Should-Have feature does not support by Aragon.

The first time the DSS suggested infeasible solutions; hence we had to relax part of the hard constraints (Must-Have and Won't-Have features) and converted them into soft constraints (Should-Have and Could-Have). for instance, the case study participants identified ``Intellectual Property'' feature as Must Have feature.
However, we had to convert it into a Should-Have feature as a soft constraint. Moreover, the ``Membership management'' feature had been identified as a Won't-Have feature by case study participants, and we converted it into None(without prioritization).

\section{Analysis of the Results}\label{AnalysisResults}

The DSS suggests that \textit{Colony}, \textit{Aragon}, and \textit{DAOStack} can be feasible solutions for all three case studies (see Table~\ref{tbl:tableSolutions}), which means that these DAO platforms support all of the features with \textit{Must-have} priority. It makes sense as these DAO platforms are in the top-5 list of popular solutions in the market (see Table~\ref{fig:NFP_DAO}); moreover, their maturity levels are relatively high, as they support most of the DAO features that we have considered in this study (see Tables~\ref{fig:BFA1} and~\ref{fig:BFA2}).

Scalability and upgradability of the DAO platforms were two key quality concerns of the case study participants (see Table~\ref{tbl:MoSCoW}) so that they considered at least one of the top-5 DAO platforms as their potential solutions. Table~\ref{tbl:tableSolutions} represents that the DSS can come up with more feasible DAO platforms than human experts (For instance, SecureSECO case study).

Table~\ref{tbl:MoSCoW} shows that supporting  \textit{Infrastructure decentralization}, \textit{On-chain}, \textit{Upgradeable contract}, \textit{Token-based voting}, \textit{Transparency portal}, \textit{Funds allocation}, \textit{Scalability}, \textit{Upgradability}, \textit{Reputation-based voting}, \textit{Governance upgrade}, \textit{Extensibility}, \textit{Permissionless}, \textit{Shared Resource}, and \textit{Proposals} were DAO features that all of the case studies assigned priorities to them and defined them as their DAO feature requirements. All of the case study participants somehow declared that the upgradeable smart contract is essential as allows us to iteratively add new features to our DAO, or fix any bugs that may find in production. 

It is not surprising that infrastructure decentralization and on-chain governance were prioritized as two essential features for all case studies,
as these two features are crucial in a DAO. One of the case study participants mentioned that with using infrastructure decentralization, there is no single point of failure; every department has the internal infrastructure to handle, analyze and manage data. Thus they are not reliant on a single central server to handle all the processes.

Another case study participants about on-chain governance mentioned 
the main advantage of on-chain governance is the codification of rules that govern the entire network and can be known by all participants in the network. Also, they mentioned that On-chain governance has several benefits over its informal counterpart, including a decentralized decision-making process, binding code changes, transparency, quicker consensus, fewer malicious hard forks. 

Table \ref{tbl:tableSolutions} shows that the case study participants who indicated the feature requirement with more confidence were advised a limited set of alternative solutions. Hence, the higher number of hard-constrained feature requirements (Must-Have) on unique programming language features leads to fewer alternative solutions.

For instance, dOrg prioritized their feature requirements according to their current main solutions (DAOStack and Aragon ), so they have assigned Must-Have priority to the particular features, such as supporting Infrastructure decentralization and Reputation-based voting. In other words, their feature requirements were biased to the features that their shortlist of DAO platforms supported them.

The results show that flexibility on the feature requirements leads to a higher number of alternative solutions. For instance, the DSS suggested a broader list of alternative solutions to the case study participants at \textit{SecureSECO} as they did not emphasize particular feature requirements and defined more soft-constrained (Should-Have and Could-Have) features. 

\section{Discussion}\label{Discussion}

The validity metric is defined as the degree to which an artifact works correctly. There are two ways to measure validity: (1) the results of the DSS compared to the predefined case-study participant shortlist of potentially feasible DAO platforms, and (2) according to the domain experts' opinion. 

Concerning effectiveness, the case study participants stated that the updated and validated version of the decision model is useful and valuable in finding the shortlist of feasible DAO platforms. Moreover, the DSS reduces the time and cost of the decision-making process. The case study participants expressed that the DSS enabled them to meet more detailed DAO feature requirements. Furthermore, they were surprised to find their primary concerns, especially when different experts' opinions are combined.

\subsection{Case Studies}

We conducted a set of interviews with the experts at three case study organizations and asked them to indicate their feature requirements based on the MoSCoW prioritization technique. If the DSS suggests infeasible solutions, we need to relax part of the hard constraints (Must-Have and Won't-Have features) and convert them into soft constraints (or Should-Have and Could-Have). Sometimes, the case study participants misunderstood the meaning of the Won't-Have priority. So, they use it to indicate what they do not care about. 

If the DSS suggests infeasible solutions, we need to relax part of the hard constraints (Must-Have and Won't-Have features) and convert them into soft constraints (or Should-Have and Could-Have). Sometimes, the case study participants misunderstood the meaning of the Won't-Have priority. So, they use it to indicate what they do not care about. 
We faced the same problem when we tried to define the feature requirements of case studies on the DSS. So, we have relaxed the constraints to come up with some feasible solutions on the DSS. 

Hence, we have ranked the feature requirements based on the number of DAO platforms that they support. We can identify the features that lead to infeasible solutions if they are prioritized as Must-Have or Won't-Have. Then, We have changed the most vulnerable features from Must-Have to Should-Have or Won't-Have to None (without priorities). We have done it iteratively until We find at least a feasible solution.(See table ~\ref{tbl:MoSCoW})

\subsection{Expert Interviews}

We did not use formal coding for the analysis of the interviews and the literature. What we did do, however, could be termed incremental concept development. During the literature study and interviews, concepts were identified that were relevant. Candidate qualities and features were identified, defined, and fine-tuned with the interviewees and later confirmed by asking the interviewees for a post-analysis of the interview and literature results. While this did not constitute formal coding, we did mark concepts related to the domain, came up with the literature study, and came up with the interviews. Secondly, these concepts were incrementally fine-tuned until an agreement was reached with the interviewees \cite{farshidi2020model}.

One of the experts asserted that smart contracts define a decentralized autonomous organization. However, a good organization also needs liquid funds. It needs to make good decisions and communicate with all instances. There is no management within a DAO, only decision-making capabilities that are executed by a code distributed across thousands of computers. Hence, smart contracts play an essential role in the DAO.

The experts expressed that technical vulnerabilities of DAOs include cybersecurity, voting procedure, and voter manipulation. They mentioned that the immutable nature of blockchain ledgers could also make the DAO vulnerable to attacks because it is so difficult to alter the essential construction of the DAO if a bug in the code appears.
So almost all of the experts mentioned that considering the security issues on DAO is crucial.

\subsection{The Decision Model}

The case study participants confirm that the updated and validated version of the DSS is helpful and valuable in finding the shortlist of feasible solutions. Finally, it decreases the time and cost of the decision-making process. Our website\footnote{https://dss-mcdm.com} is up and running to keep the decision support system's knowledge base up-to-date and valid. The supported DAO platform features are going to change due to technological advances. As such, the decision model must be updated regularly. We envision a community of users of the DSS who maintain and curate the system's knowledge and consider building such a community as future work.

Decision support systems can be employed to make decisions quicker and more efficiently; however, they suffer from adoption problems~\cite{donzelli2006decision}. A DSS supports rational decision making by recommending alternative solutions basis the objectivity. Although limited rationality plays a crucial role in a decision-making process, subjectivity should not be discarded. A DSS promotes objectivity and dismisses subjectivity, which can have a drastic consequence on the decisions' reliability.

We believe that the theoretical contribution and the answer of the main research question (see section~\ref{DAO_ResearchQuestion}) of this study is a decision model that can be used to make informed decisions in software production, and models from software engineering, such as the ISO standard quality model and the MoSCoW prioritization technique, are fundamental building blocks in such decisions. Researchers can replace the ISO standard quality model with more specific quality attributes to customize the decision model. Although we employ the MoSCoW prioritization technique to simplify the understanding and manage priorities, other researchers can employ other types of prioritization techniques to define the feature requirements.

Researchers can more rapidly evaluate DAO platforms in the market by using the knowledge available through the decision model, and also They can add more platforms or features to the decision model systematically according to the presented guideline, employ the reusable knowledge (presented in Tables \ref{fig:BFA1}, \ref{fig:BFA2}, \ref{fig:NFP_DAO}, and \ref{tbl:MoSCoW}) to develop new concepts and solutions for future challenges.

\subsection{Limitations and Threats to Validity}
The validity assessment is an essential part of any empirical study. Validity discussions typically involve Construct Validity, Internal Validity, External Validity, and Conclusion Validity. 

\noindent\textbf{Construct validity} refers to whether an accurate operational measure or test has been used for the concepts being studied. In literature, decision-making is typically defined as a process or a set of ordered activities concerning stages of problem identifying, data collection, defining alternatives, selecting a shortlist of alternatives as feasible solutions with the ranked preferences~\cite{fitzgerald2017differences,kaufmann2012rationality}. To mitigate the threats to the construct validity, we followed the MCDM theory and the six-step of a decision-making process~\cite{Majumder2015} to build the decision model for the DAO platform selection problem. Moreover, we employed document analysis and expert interviews as two different knowledge acquisition techniques to capture knowledge regarding DAO platforms. Additionally, the DSS and the decision model have been evaluated through three real-world case studies at three different real-world enterprises in the Netherlands, United States, and Iran.

\noindent\textbf{Internal validity} attempts to verify claims about the cause-effect relationships within the context of a study. In other words, it determines whether the study is sound or not. To mitigate the threats to the decision model's internal validity, we define DSS success when it, in part, aligns with the case study participants' shortlist and when it provides new suggestions that are identified as being of interest to the case study participants.  Emphasis on the case study participants' opinion as a measurement instrument is risky, as they may not have sufficient knowledge to make a valid judgment. We counter this risk by conducting more than one case study, assuming that the case study participants are handling their interest and applying the DSS to other problem domains, where we find similar results~\cite{farshidiCSP,Siamak2018DBMS,farshidi2018DSS,FarshidiBlockchain2019,FarshidiToolPaper2020,FarshidiSLR_Arxiv,farshidi2021decision}. 

\noindent\textbf{External validity} concerns the domain to which the research findings can be generalized. External validity is sometimes used interchangeably with generalizability (feasibility of applying the results to other research settings). We evaluated the decision model in the context of Dutch enterprises. To mitigate threats to the research's external validity, we captured knowledge from different sources of knowledge without any regional limitations to define the constructs and build the decision model. Accordingly, we hypothesize that the decision model can be generalized to all decentralized companies and organizations that face uncertainty in the DAO platform selection problem.
Another question is whether the framework and the DSS can be applied to other problem domains as well. The problem domains~\cite{farshidiCSP,Siamak2018DBMS,FarshidiBlockchain2019,FarshidiSLR_Arxiv} were selected opportunistically and pragmatically, but we are convinced that there are still many decision problems to which the framework and the DSS can be applied. The categories of problems to which the framework and the DSS can be applied successfully can be summed up as follows: (1) the problem regards a technology decision in system design with long-lasting consequences, (2) there is copious scientific, industry, and informal knowledge publicly available to software engineers, and (3) the (team of) software engineer(s) is not knowledgeable in the field but very knowledgeable about the system requirements.

\noindent\textbf{Conclusion validity} verifies whether the methods of a study such as the data collection method can be reproduced, with similar results. We captured knowledge systematically from the sources of knowledge following the MCDM framework~\cite{farshidiCSP}. The accuracy of the extracted knowledge was guaranteed through the protocols that were developed to define the knowledge extraction strategy and format (See appendix~\ref{InterviewProtocols}).  A review protocol was proposed and applied by multiple research assistants, including bachelor and master students, to mitigate the threats to the research's conclusion validity. By following the framework and the protocols, we keep consistency in the knowledge extraction process and check whether the acquired knowledge addresses the research questions. Moreover, we crosschecked the captured knowledge to assess the quality of the results, and we had at least two assistants extracting data independently.

\nopagebreak
\afterpage{

\begin{landscape}
\begin{table}[!ht]
\scriptsize
\centering
\caption{This table compares a subset of selected studies from the literature that addresses the DAO platform selection problem. The first two columns indicate the selected study (Study) and the publication year (Year).The next column (MCDM) denotes whether the corresponding decision-making technique is an MCDM approach. The next three columns Quality Attribute(QA) determines the type of quality attributes, the data collection type (Data Col.) of the corresponding selected studies, respectively. The sixth column (Approach) indicates the decision-making approach that the studies have employed to address the DAO platform selection problem. the seventh and eighth column indicate research methods (R. Method) (including Expert Interview, Document Analysis, Design Science and Case Study) and publication type (including Research Paper, Dissertation, Chapter and Report) of the corresponding selected studies, respectively. The ninth and tenth and eleventh and twelfth columns(\#F and \#QA and \#C and \#A) signify the number of features and quality attributes and criteria and alternatives considered in the selected studies. The next three columns indicate the numbers of features (\#CF), common quality attributes (\#CQA), and alternatives (\#CA) of this study (the first row) with the selected studies. The last column (Cov.) shows the percentage of the coverage of the considered criteria (quality attributes and features).}
\label{DAO_tableLiterature}
\scriptsize
\centering

\begin{tabular}{|l|r|l|l|l|l|l|l|r|r|r|r|r|r|r|r|}
\hline
\textbf{Study} & \multicolumn{1}{l|}{\textbf{Year}} & \textbf{MCDM} & \textbf{QA} & \textbf{Data Col.} & \textbf{Approach} & \textbf{R. Method} & \textbf{publication type} & \multicolumn{1}{l|}{\textbf{\#F}} & \multicolumn{1}{l|}{\textbf{\#QA}} & \multicolumn{1}{l|}{\textbf{\#C}} & \multicolumn{1}{l|}{\textbf{\#A}} & \multicolumn{1}{l|}{\textbf{\#CF}} & \multicolumn{1}{l|}{\textbf{\#CQA}} & \multicolumn{1}{l|}{\textbf{\#CA}} & \multicolumn{1}{l|}{\textbf{Cov.)}} \\ \hline\hline
Our study & 2021& Yes& \begin{tabular}[c]{@{}l@{}}ISO/IEC 25010\\ Ext. ISO/IEC 9126\end{tabular} & Mixed & DSS& \begin{tabular}[c]{@{}l@{}}Case Study\\ Expert interview\\ Document Analysis\end{tabular} & Research paper & 82 & 46& 128& 28 & 82& 46 & 28& 100\% \\ \hline\hline
\cite{valiente2020evaluating} & 2020& No & Domain specific & Qualitative & Benchmarking & Case Study& Research paper & 9& 2 & 11 & 3& 8 & 2& 3 & 91\%\\ \hline
\cite{liu2020technology}& 2020& No & Domain specific & Qualitative & Benchmarking & Document Analysis & Research paper & 4& 0 & 4& 3& 4 & 0& 3 & 100\% \\ \hline
\cite{ziolkowski2020exploring} & 2020& No & N/A & Qualitative & Benchmarking & \begin{tabular}[c]{@{}l@{}}Document Analysis\\ Expert Interview\end{tabular}& Research paper & 3& 0 & 3& 3& 2 & 0& 2 & 67\%\\ \hline
\cite{lee2020decision} & 2020& Yes& Domain specific & Mixed & \begin{tabular}[c]{@{}l@{}}AHP \\ ANP \\ NGT\end{tabular}& \begin{tabular}[c]{@{}l@{}}Document Analysis\\ Expert Interview\end{tabular}& Research paper & 4& 2 & 6& 3& 4 & 2& 1 & 100\% \\ \hline
\cite{skarzauskiene2020developing} & 2020& No & Domain specific & Qualitative & Benchmarking & Document Analysis & Research paper & 13 & 5 & 18 & 1& 12& 4& 0 & 89\%\\ \hline
\cite{sims2019blockchain} & 2020& No & Domain specific & Quantitative & Benchmarking & Document Analysis & Research paper & 6& 0 & 6& 6& 6 & 0& 2 & 100\% \\ \hline
\cite{faqir2020scalable} & 2020& No & Domain specific & Quantitative& Statistical analysis & Document Analysis & Research paper & 4& 1 & 4& 4 & 4 & 1& 4 & 25\%\\ \hline
\cite{rikken2019governance} & 2019& No & Domain specific & Qualitative & Benchmarking & Expert interview& Research paper & 3& 0 & 3& 1& 2 & 0& 0 & 67\%\\ \hline
\cite{el2020overview} & 2018& No & Domain specific & Quantitative& Statistical analysis & Document Analysis & Research paper & 7& 0 & 7& 4& 1 & 0& 4 & 14\%\\ \hline
\cite{dubeinformation} & 2018& No & Domain specific & Qualitative & Benchmarking & Document Analysis & Research paper & 9& 0 & 9& 1& 2 & 0& 1 & 22\%\\ \hline
\cite{valienteresults} & 2017& No & Domain specific & Qualitative & Benchmarking & \begin{tabular}[c]{@{}l@{}}Document Analysis\\ Case Study\end{tabular}& Research paper & 9& 0 & 9& 3& 8 & 0& 3 & 89\%\\ \hline
\cite{hsieh2018rise}& 2018& Yes& Domain specific & Mixed & Fuzzy logic& Document Analysis & Dissertation & 9& 4 & 13 & 9& 5 & 3& 0 & 62\%\\ \hline
\cite{dupont2017experiments} & 2017& No & Domain specific & Qualitative & Benchmarking & \begin{tabular}[c]{@{}l@{}}Document Analysis\\ Grounded theory\end{tabular} & Chapter& 2& 0 & 2& 1& 2 & 0& 1 & 100\% \\ \hline
\citeONLINE{tan_2021:Online} & 2020& No & Domain specific & Qualitative & Benchmarking & Document Analysis & Report & 2& 0 & 2& 8& 2 & 0& 2 & 100\% \\ \hline
\cite{Felix_Machart} & 2020& No & Domain specific & Qualitative & Benchmarking & Document Analysis & Report & 8& 0 & 8& 9& 8 & 0& 8 & 100\% \\ \hline
\citeONLINE{DAO_Overview} & 2020& No & Domain specific & Qualitative & Benchmarking & Document Analysis & Report & 9& 0 & 9& 8& 5 & 0& 4 & 56\%\\ \hline
\citeONLINE{George_Samman} & 2020& No & Domain specific & Qualitative & Benchmarking & Document Analysis & Report & 7& 0 & 7& 3& 7 & 0& 10& 100\% \\ \hline
\citeONLINE{Theoryan8:online} & 2019& No & Domain specific & Qualitative & Benchmarking & Document Analysis & Report & 24 & 0 & 24 & 6& 24& 0& 6 & 100\% \\ \hline
\citeONLINE{weller_2019:Online} & 2019& No & Domain specific & Qualitative & Benchmarking & Document Analysis & Report & 8& 0 & 8& 1& 5 & 0& 1 & 63\%\\ \hline
\citeONLINE{noauthor_aragon_nodate:Online} & 2019& No & Domain specific & Qualitative & Benchmarking & Document Analysis & Report & 3& 0 & 3& 4& 2 & 0& 4 & 67\%\\ \hline
& \multicolumn{1}{l|}{} && & && && \multicolumn{1}{l|}{}& \multicolumn{1}{l|}{} & \multicolumn{1}{l|}{}& \multicolumn{1}{l|}{}& \multicolumn{1}{l|}{} & \multicolumn{1}{l|}{}& \multicolumn{1}{l|}{} & 79.24\% \\ \hline
\end{tabular}
\end{table}
\end{landscape}
\vfill
}

\section{Related Work} \label{Relatedwork}

In this study, Snowballing was applied as the primary method to investigate the existing literature regarding techniques that address the DAO platform selection problem. Table \ref{DAO_tableLiterature} summarizes a subset of selected studies that discuss the problem. As aforementioned, the last column (\textit{Cov.}) of Table~\ref{DAO_tableLiterature} indicates the percentage of the coverage of the considered criteria within the selected studies. On average, 79.24\% of those criteria are already considered in this study. In other words, the decision model contains a significant number of criteria, including features and quality attributes, that have been mentioned in the literature.

\subsection{Benchmarking and Statistical Analysis}
Some studies employed Benchmarking and Statistical Analysis to evaluate and compare a collection of DAO platforms against each other in literature. For instance, Valiente et al.~\cite{valiente2020evaluating} perform an analytical comparison of three DAO software frameworks: Aragon, DAOstack, and Colony. They focus on their current functionalities for building DAOs, and they present a case study using the Aragon framework. They are performed with the case study of a sample DAO that supports researchers participating in a typical project to manage the different tasks they have to carry out. 

Liu et al. review the most recent research activities on academic and engineering scenarios, including governance problems and solutions, typical DAO technologies, and related areas. They perform such an overview by identifying and classifying the most valuable proposals and perspectives related to the combination of DAO and blockchain technologies~\cite{liu2020technology}. 

Ziolkowski et al.~\cite{ziolkowski2020exploring} explore multiple case study consisting of three famous DAOs, Aragon, Tezos, and DFINITY. This study introduced each case by depicting the DAOs' organizational and technological structure and brought forward concepts. Second, They have created an understanding of how these days are governed by examining their governance systems in terms of applied/envisioned coordination, control, and incentive mechanisms. As they study fewer DAO features and DAO platforms, our work could be considered more comprehensive.  They studied only three DAO platforms and four DAO features against our study that we have considered 82 DAO features and 28 DAO platforms.

Faqir et al.~\cite{el2020overview} introduce the concept of DAO and review the primary software platforms that offer DAO creation as a service, which simplifies the use of DAOs to non-blockchain experts, namely: Aragon, DAOstack, DAOhaus, and Colony. These platforms are compared by showing their key features. Finally, the authors will review the available visualization tools for DAOs. They introduced their open-source tool to plot DAOs activity and to analyze.

Studies based on benchmarking and statistical analysis are typically time-consuming approaches and mainly applicable to a limited set of alternatives and criteria, as they require a thorough knowledge of DAO platforms and concepts. Decision-making based on such analysis can be challenging as decision-makers cannot assess all their requirements and preferences simultaneously, especially when the number of requirements and alternatives is significantly high. Furthermore, benchmarking and statistical analysis are likely to become outdated soon and should be kept up to date continuously, which involves a high-cost process. 

\subsection{MCDM approaches}
Selecting the best fitting DAO platform is a decision-making process that evaluates several alternatives and criteria. The selected DAO platform should address the concerns and priorities of the decision-makers. Conversely to MCDM approaches, studies based on \textit{Benchmarking} and \textit{Statistical Analysis} principally offer generic results and comparisons and do not consider individual decision-maker needs and preferences.

The tools and techniques based on MCDM are mathematical decision models aggregating criteria, points of view, or features~\cite{floudas2008encyclopedia}. Support is a fundamental concept in MCDM, indicating decision models are not developed following a process where the decision maker's role is passive~\cite{dvovrak2018affordance}. Alternatively, an iterative process is applied to analyze decision-makers' priorities and describe them as consistently as possible in a suitable decision model. This iterative and interactive modeling procedure forms the underlying principle of decision support tendency of MCDM, and it is one of the main distinguishing characteristics of the MCDM as opposed to statistical and optimization decision-making approaches~\cite{gil2013handbook}.

A variety of MCDM approaches have been introduced by researchers recently. A subset of selected MCDM methods is presented as follows: The \textit{Analytic Hierarchy Process (AHP)} is a structured and well-known method for organizing and analyzing MCDM problems based on mathematics and psychology. The analytic network process (ANP) is structured on the same basis as AHP; however, it differs from AHP in two ways. (1) ANP does not assume that the alternatives and criteria are independent. The feedback mechanism handles their potential dependencies. (2) ANP has a network structure that forms subnetworks and submodels. The nominal group technique (NGT) is a group decision-making process including problem identification, solution generation, and decision-making. 

Yosep et al.~\cite{lee2020decision} examined the decision-making process and tools applicable to a decentralized autonomous organization. This paper studied a decision-making process that features iteration, visualization, and applicability to DAO with six steps in total and a decision-making tool based on this paper's process. Traditional methods such as AHP, ANP, and NGT have been studied in this paper. 

Fuzzy logic is an approach to computing based on \textit{degrees of truth} rather than the usual Boolean logic. Valiente et al.~\cite{hsieh2018rise} considered a set of DAO platforms for finding the right option for a case study. They performed fuzzy logic in their analysis as a tool for decision-making. Additionally, the authors explained the conceptualization of DAOs and the defining features of coordination mechanisms within DAOs.

The majority of the MCDM techniques in literature define domain-specific quality attributes to evaluate the alternatives. Such studies are mainly appropriate for specific case studies. Furthermore, the results of MCDM approaches are valid for a specified period; therefore, the results of such studies will be outdated by DAO platform advances. Note that, in our proposal, this is also a challenge, and we propose a solution for keeping the knowledge base up to date in section~\ref{Discussion}.

\subsection{Strengths and liabilities of the decision model}

Determining the best DAO platform for an organization is a decision-making process that involves evaluating various alternatives and criteria. Hence, the selected platform should address the concerns and priorities of the decision-makers. Studies based on 'Benchmarking' and 'Statistical Analysis', in contrast to MCDM techniques, primarily provide general results and comparisons and do not consider individual decision-maker requirements and preferences. Benchmarking and Statistical Analysis (SA) methods are often time-consuming and only apply to a small number of alternatives and criteria. Furthermore, benchmarking and statistical analysis are likely to become obsolete quickly and must be maintained up to date regularly, which is a costly operation. 

Researchers have presented a range of MCDM approaches in the literature. To evaluate the alternatives, the majority of MCDM approaches establish domain-specific quality attributes. Some approaches, such as Fuzzy and AHP, are not scalable; the evaluation process needs to be performed if the list of alternatives or criteria is changed. Accordingly, these approaches are expensive and only apply to a limited set of criteria and alternatives.

This study has considered 82 criteria and 28 alternatives to building a decision model for the DAO platform selection problem. MCDM approach is an evolvable and expandable approach that divides down the decision-making process into four maintainable phases. 

In contrast to the methods mentioned above, the cost of creating, evaluating, and applying the proposed decision model is not penalized exponentially by the number of criteria and alternatives~\cite{Siamak2018DBMS} . Additionally, we introduce several parameters to estimate the values of non-Boolean criteria, such as the maturity level and market popularity of the DAO platforms. The proposed decision model addresses the most important aspects of knowledge management, such as knowledge capture, sharing, and maintenance. Furthermore, it uses the ISO/IEC 25010~\cite{iso2011iec25010} as a standard set of quality attributes. This quality standard is a domain-independent software quality model and provides reference points by defining a top-down standard quality model for software systems. 

Recently, we have built six decision models based on the framework to model the selection of database management systems~\cite{Siamak2018DBMS}, cloud service providers~\cite{farshidiCSP}, blockchain platforms~\cite{FarshidiBlockchain2019}, software architecture patterns~\cite{FarshidiSLR_Arxiv}, model-driven platforms~\cite{farshidi2020model}, and programming languages~\cite{farshidi2021decision}. These case studies were conducted to evaluate the DSS's effectiveness and usefulness in addressing MCDM problems. The results confirmed that the DSS performed well to solve the mentioned problems in software production. We believe that the framework can be employed as a guideline to build decision models for MCDM problems in software production.

\section{Conclusion and Future Work}\label{CONCLUSION}

In this study, the DAO platform selection process is modeled as a multi-criteria decision-making problem that deals with evaluating a set of alternatives and considering a set of decision criteria~\cite{triantaphyllou1998multi}. Moreover, we presented a decision model for the DAO platform selection problem based on the technology selection framework~\cite{farshidiCSP}. The approach provides knowledge about DAO platforms to support uninformed decision-makers while contributing a sound decision model to knowledgeable decision-makers. The framework incorporates deeply embedded requirements engineering concepts (such as the ISO software quality standards and the MoSCoW prioritization technique) to develop the decision model. 

The scientific contributions of this work are threefold. First, the methodical collection of features from a multitude of resources supports researchers who need a comprehensive overview of DAOs and their features. Secondly, we prove that the MCDM approach and its supporting decision support system are valuable in new contexts for technology selection. Finally, we show that case studies are an excellent research method for evaluating designed artifacts, such as the MCDM framework.

We conducted three industry case studies to evaluate the decision model's usefulness and effectiveness to address the decision problem. We find that while organizations are typically tied to particular ecosystems by extraneous factors, they can benefit significantly from our DSS by evaluating their decisions, exploring more potential alternative solutions, and analyzing an extensive list of features.

The case studies show that this article's decision model also provides a foundation for future work on MCDM problems. We intend to build trustworthy decision models to address the \textit{Consensus Algorithm} selection problem and the \textit{self-sovereign identity framework} selection problem as our (near) future work.

\balance
\bibliographystyle{abbrv}
\bibliography{references}

\bibliographystyleONLINE{abbrv}
\bibliographyONLINE{references}

\appendix

\section{Expert Interview Protocols}\label{InterviewProtocols}

\fbox{\begin{minipage}{23.5em}
\scriptsize
\textbf{[Evaluation of the DAO features and platforms]}\newline\newline

\noindent\textbf{Step 1-} A brief description of the project, the decision model, the DSS, and the main goal of the interview.   \newline

\noindent\textbf{Step 2-} Introductory questions: \newline
\textit{- What do you understand about DAO?}\newline
\textit{- Which features of DAO do you familiar with?}\newline
\textit{- Which DAO Platform are you familiar with?}\newline
\textit{- How long have you worked on DAO platforms?}\newline

\noindent\textbf{Step 3-} Decision-making questions: \newline
\textit{- Why do you need a DAO for your company?}\newline
\textit{- How do a decentralized organization typically select DAO platforms?}\newline
\textit{- What are the essential features from your perspective for selecting the best fitting DAO platform?}\newline
\textit{- Which DAO platforms are typically considered as alternative solutions by decentralized organizations?}\newline

\noindent\textbf{Step 4-} Evaluation of the sets of DAO platform / features:\newline
\textit{- What do you think about these DAO platform / features?}\newline
\textit{- Which DAO platform / features should be excluded from the list?}\newline
\textit{- Which DAO platform /features should be added to the list?}\newline

\noindent\textbf{Step 5-} Closing\newline
\textit{- What do you think about our work?}\newline
\textit{- May we contact you if we have any further questions? }\newline
\textit{- Can we use the name of your company in the scientific paper, or do you prefer an anonymous name?}\newline
\textit{- Can we use your name in the scientific paper, or do you prefer an anonymous name? }\newline
\textit{- Do you have any questions?}\newline
\end{minipage}}
\newline\newline\newline
\noindent\fbox{\begin{minipage}{23.5em}
\scriptsize
\textbf{[Mapping between the DAO features and the quality attributes]}\newline\newline

\noindent\textbf{Step 1-} A brief description of the project, the decision model, the DSS, and the main goal of the interview.  \newline

\noindent\textbf{Step 2-} Introductory questions: \newline
\textit{- What do you understand about DAO?}\newline
\textit{- Which features of DAO do you familiar with?}\newline
\textit{- Are you familiar with the ISO/IEC quality models?}\newline

\noindent\textbf{Step 3-} Mapping between the DAO features and the quality attributes: (Note: this step will be repeated for all of the features and quality attributes).\newline
\textit{- Does the DAO feature [X] have a positive impact on the quality attribute [Y]? For instance, if a DAO platform supports \textit{Conviction Voting} means that it has positive impacts on \textit{Functional appropriateness} and \textit{Stability}.}\newline

\noindent\textbf{Step 4-} Closing\newline
\textit{- What do you think about our work?}\newline
\textit{- May we contact you if we have any further questions? }\newline
\textit{- Can we use the name of your company in the scientific paper, or do you prefer an anonymous name?}\newline
\textit{- Can we use your name in the scientific paper, or do you prefer an anonymous name? }\newline
\textit{- Do you have any questions?}\newline
\end{minipage}}



\begin{table*}[!ht]
  \caption{This table shows the first part of the Boolean Features ($Feature^{B}$), DAO Platforms ($Platforms$), and the ``BFP'' mapping. Note, 1s on each row indicates that the corresponding platforms support the DAO feature of that row, and 0s signify the corresponding platforms do not support that feature, or we did not find any strong evidence of their supports based on the documentation analysis. Moreover, the rows in black indicate the categories of the features, and the rows in blue show the features, and the rows below them are their sub-features.}
  \centering
   \includegraphics[trim=0 0 0 0,clip,width=0.85\textwidth]{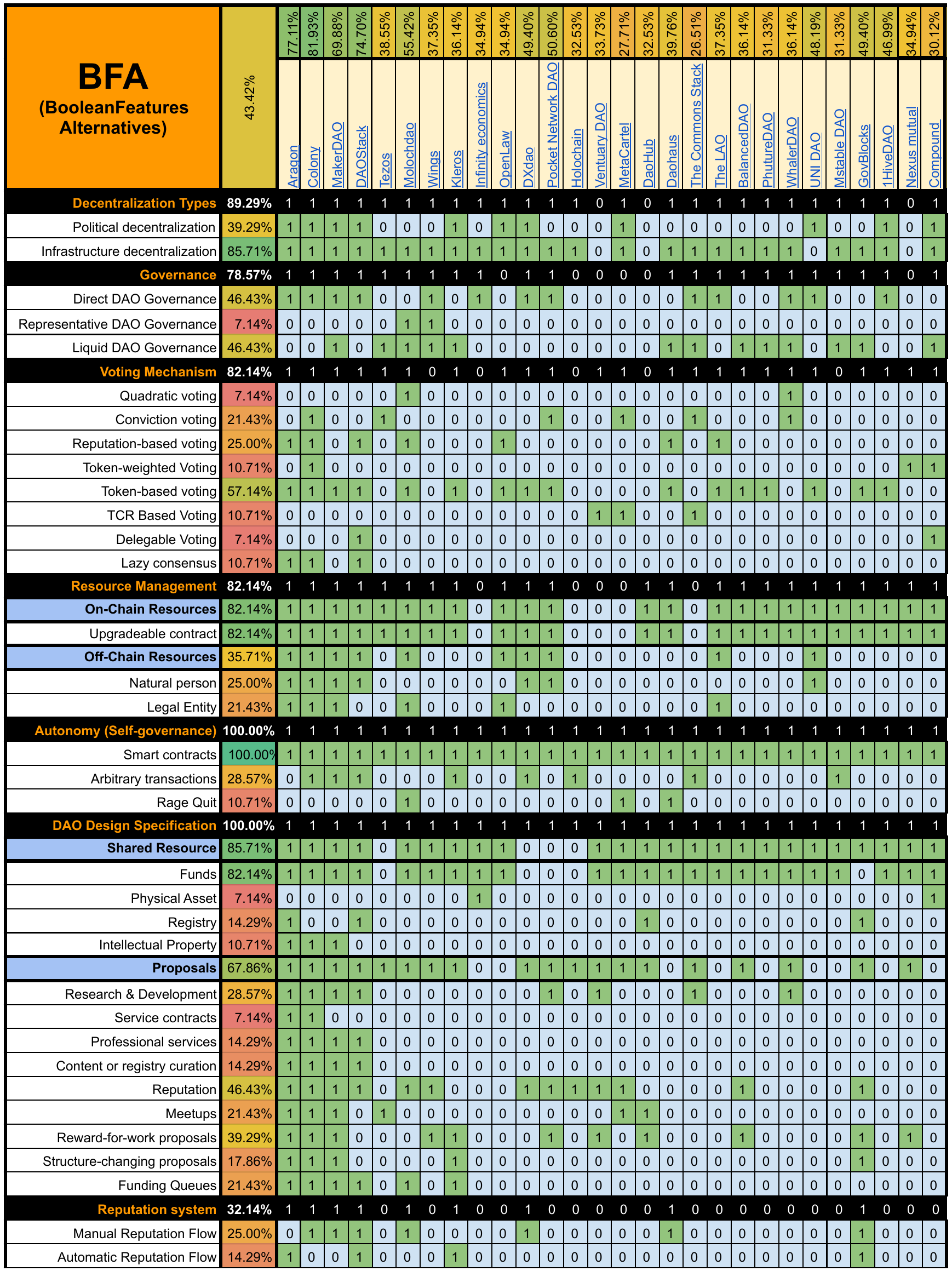}
  \label{fig:BFA1}
\end{table*}

\begin{table*}[!ht]
  \caption{This table shows the second part of the Boolean Features ($Feature^{B}$), DAO Platforms ($Platforms$), and the ``BFP'' mapping. Note, 1s on each row indicates that the corresponding platforms support the DAO feature of that row, and 0s signify the corresponding platforms do not support that feature, or we did not find any strong evidence of their supports based on the documentation analysis. Moreover, the rows in black indicate the categories of the features, and the rows in blue show the features, and the rows below them are their sub-features. The definitions of the features are available on the data repository~\cite{farshidi2020model}.}
  \centering
   \includegraphics[trim=0 0 0 0,clip,width=0.85\textwidth]{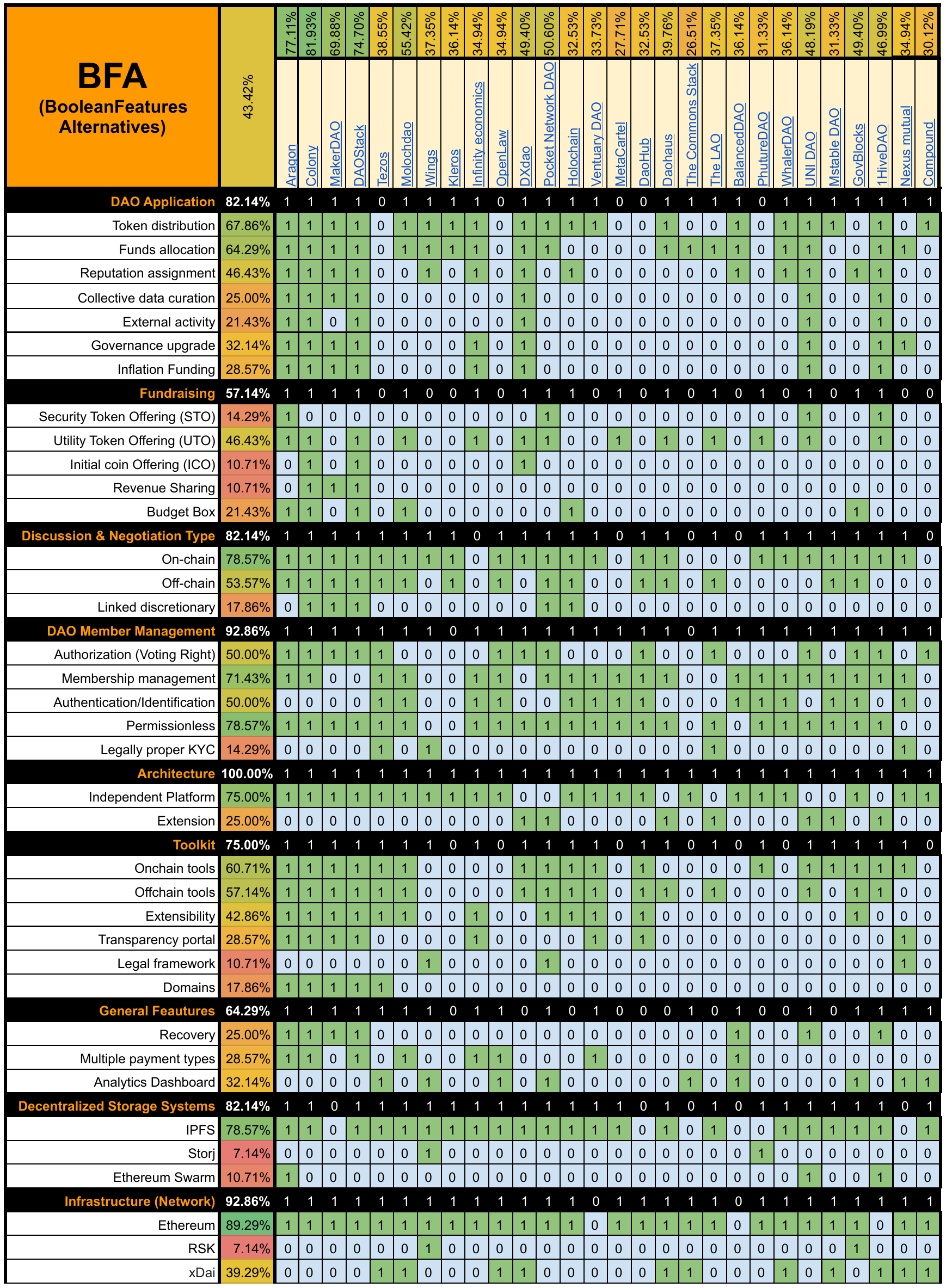}
  \label{fig:BFA2}
\end{table*}

\begin{table*}[!ht]
  \caption{This table shows the \textbf{NFP} mapping between the \textbf{N}on-Boolean DAO \textbf{F}eatures and \textbf{P}latforms. }
  \centering
   \includegraphics[trim=20 0 20 50,clip,width=0.9\textwidth]{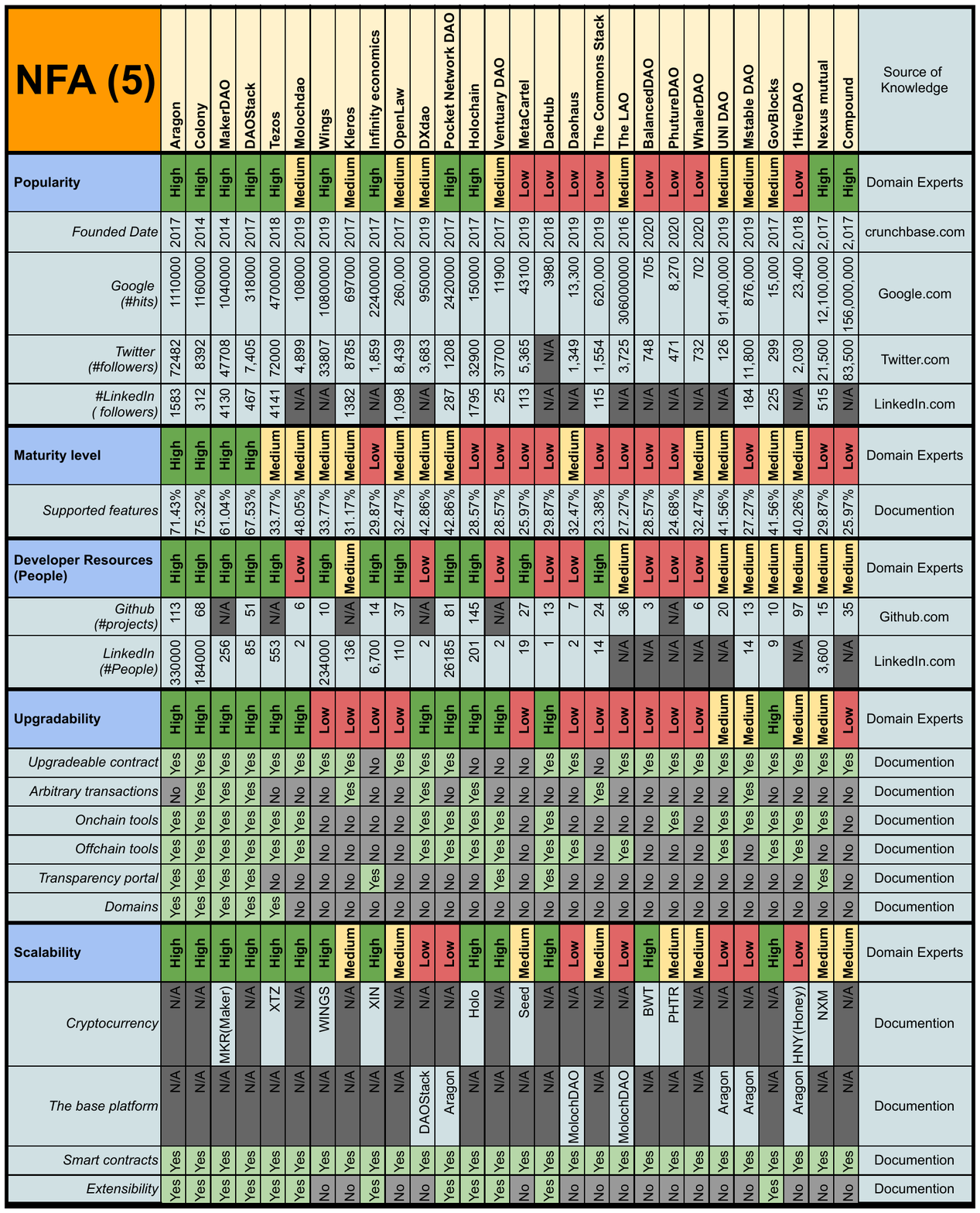}
  \label{fig:NFP_DAO}
\end{table*}

\end{document}